\pdfoutput=1
\documentclass{sig-alternate}
\usepackage{amssymb,amsmath,mathrsfs}
\usepackage{graphicx}
\usepackage[linesnumbered,ruled,lined]{algorithm2e}
\usepackage[small]{subfigure}
\usepackage[font=footnotesize,labelfont=bf]{caption} 
\usepackage{color}
\usepackage{url}
\usepackage{float}
\usepackage{multirow}
\usepackage{tabularx}
\usepackage{booktabs}
\usepackage{enumitem}

\newtheorem{definition}{Definition}

\newtheorem{hypothesis}{Hypothesis}

\DeclareMathOperator{\R}{\mathbb{R}}

\DeclareMathOperator*{\argmax}{argmax}
\newcommand*{\argmaxl}{\argmax\limits}

\def\etal{{\em et al.\/}\,}
\def\st{\mathrm{s.t.}}
\def\ie{{\sl i.e.}}
\def\eg{{\sl e.g.}}

\def\F{\mathcal{F}}
\def\G{\mathcal{G}}
\def\C{\mathcal{C}}
\def\D{\mathcal{D}}
\def\U{\mathcal{U}}
\def\P{\mathcal{P}}
\def\Ppos{\mathcal{P}^+}
\def\Q{\mathcal{Q}}
\def\Qp{\mathcal{Q}^\prime}
\def\I{\mathcal{I}}
\def\O{\mathcal{O}}
\def\H{\mathcal{H}}
\def\S{\mathcal{S}}
\def\Sp{\mathcal{S}^\prime}

\def\d'{{d^\prime}}

\newcommand{\nop}[1]{}
\newcolumntype{t}{>{\hsize=.1\hsize}c}
\newcolumntype{j}{>{\hsize=.2\hsize}X}
\newcolumntype{s}{>{\hsize=.4\hsize}X}
\newcolumntype{x}{>{\hsize=.5\hsize}X}
\newcolumntype{a}{>{\hsize=.45\hsize}X}

\newfont{\mycrnotice}{ptmr8t at 7pt}
\newfont{\myconfname}{ptmri8t at 7pt}
\let\crnotice\mycrnotice
\let\confname\myconfname
\begin{document}

\title{Comparative Document Analysis for Large Text Corpora}

\author{
\alignauthor
Xiang Ren$^{\dag}$\thanks{Work done when author was an intern at MSR.}$\quad$\ Yuanhua Lv$^{\ddagger}\quad$ Kuansan Wang$^{\ddagger}\quad$ Jiawei Han$^{\dag}$\\[0.5ex]
\affaddr{$^{\dag}$ University of Illinois at Urbana-Champaign, Urbana, IL, USA}\\[0.5ex]
\affaddr{$^{\ddagger}$ Microsoft Research, Redmond, WA, USA}\\[0.5ex]
\email{\begin{footnotesize}$^{\dag}$\{xren7, hanj\}@illinois.edu$\quad$ $^{\ddagger}$\{yuanhual, Kuansan.Wang\}@microsoft.com\end{footnotesize}}
}

\maketitle

\begin{abstract}
This paper presents a novel research problem on \textit{joint} discovery of commonalities and differences between two individual documents (or document sets), called Comparative Document Analysis (CDA). Given any pair of documents from a document collection, CDA aims to automatically identify sets of quality \textit{phrases} to summarize the commonalities of \textit{both} documents and highlight the distinctions of each \textit{with respect to the other} informatively and concisely. Our solution uses a general graph-based framework to derive novel measures on phrase \textit{semantic commonality} and \textit{pairwise distinction}, and guides the selection of sets of phrases by solving two joint optimization problems. We develop an iterative algorithm to integrate the maximization of phrase commonality or distinction measure with the learning of phrase-document semantic relevance in a mutually enhancing way. Experiments on text corpora from two different domains---scientific publications and news---demonstrate the effectiveness and robustness of the proposed method on comparing individual documents. Our case study on comparing news articles published at different dates shows the power of the proposed method on comparing document sets.
\end{abstract}

\section{Introduction}
\label{sec:intro}
Comparative text mining is concerned with identifying common and different information to distinguish text items in corpora. Instead of measuring the similarity between items or predicting sentiment rating on an item, it delves deeper into the text and focuses on extracting content units of different granularity (\eg, words, sentences) to summarize the commonalities and differences. 
With the rapid emergence of text-based data in many domains,
automatic techniques for comparative text mining have a wide range of applications including social media analysis (\eg, opinion mining on twitter~\cite{tkachenko2014generative,lu2008opinion}), business intelligence (\eg, customer review analysis~\cite{kim2009generating,jindal2006identifying}, news summarization~\cite{huang2014comparative,shen2010multi,lin2002single}), and scientific literature study (\eg, patentability search~\cite{zhang2015patentcom}).

\begin{figure}
\centering
\vspace{-0.4cm}
\includegraphics[width = 73 mm]{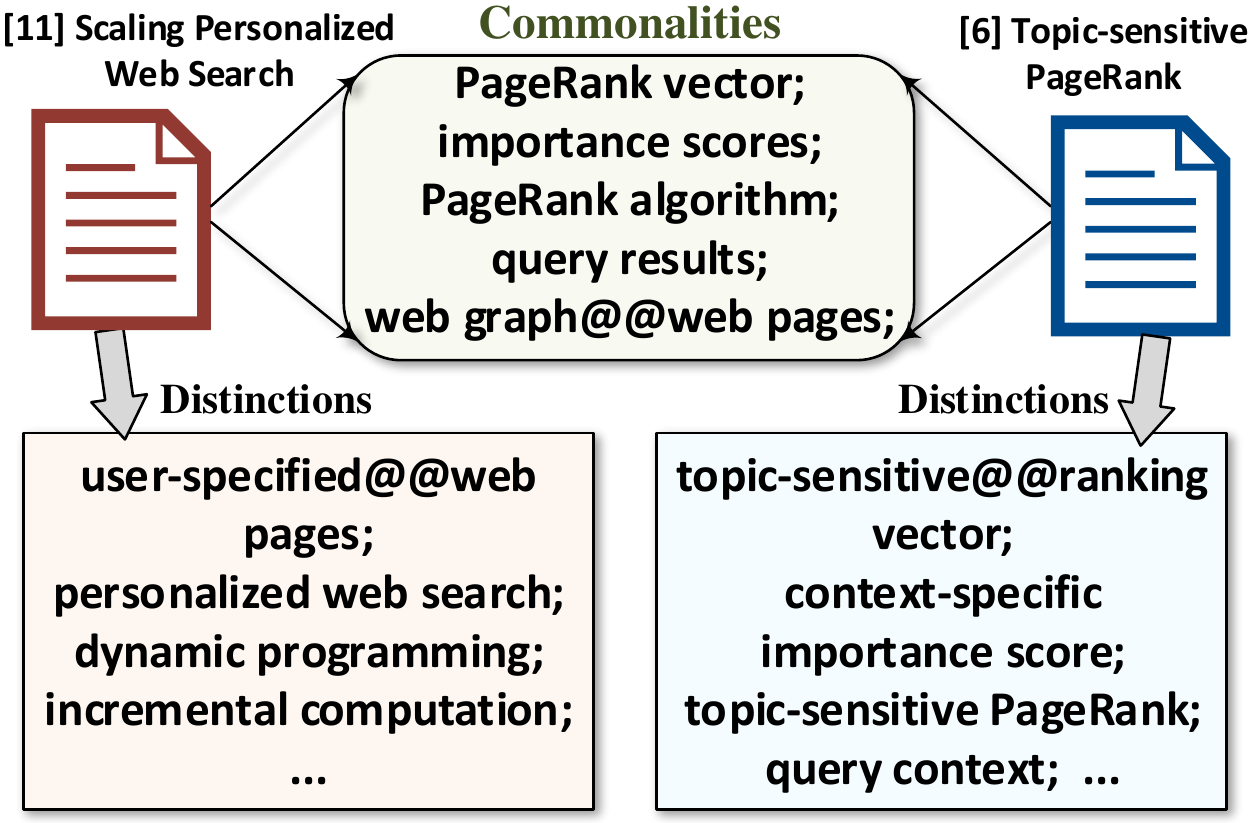}
\vspace{-0.1cm}
\caption{Example output of CDA for papers \cite{jeh2003scaling} and \cite{haveliwala2003topic}.}
\label{figure:motivated_example}
\vspace{-0.5cm}
\end{figure}

While there has been some research in comparative text mining, most of these focus on generating word-based or sentence-based summarization for sets of documents. 
Word-based summarization~\cite{mani1997summarizing,zhai2004cross} suffers from limited readability as single words are usually non-informative and bag-of-words representation does not capture the semantics of the original document well---it may not be easy for users to interpret the combined meaning of the words.
Sentence-based summarization~\cite{huang2014comparative,wang2012comparative,lerman2009contrastive}, on the other hand, may be too verbose to accurately highlight the \textit{general} commonalities and differences---users may be distracted by the irrelevant information that it contains (see Table~\ref{table:case_study_phraseVSsentence}).  
Furthermore, previous work compares two sets of documents based on the  data redundancy (\eg, word overlap) between them but the task becomes much more challenging when comparing two individual documents, since there may not exist sufficient amount of common content units between them.

\begin{figure*}
\centering
\vspace{-0.5cm}
\includegraphics[width = 172 mm]{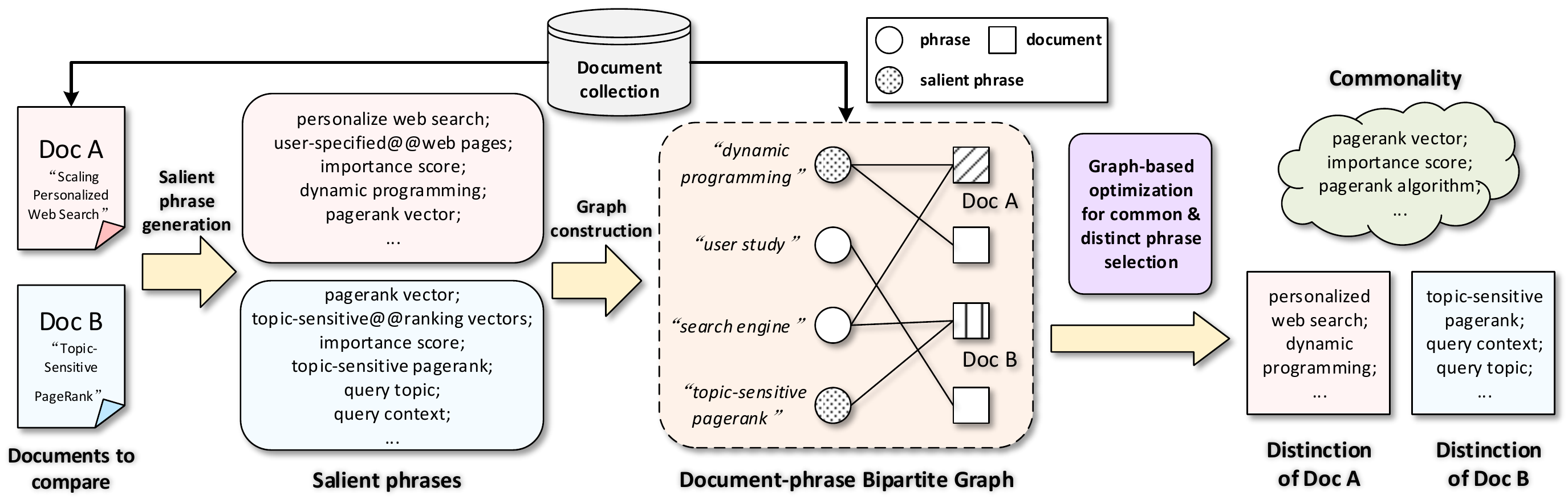}
\vspace{-0.2cm}
\caption{Overview framework of \textbf{PhraseCom}.}
\label{figure:framework_overview}
\vspace{-0.4cm}
\end{figure*}

This paper studies a novel comparative text mining problem which
leverages \textit{multi-word phrases} (\ie, minimal semantic units) to represent the common and distinct information between \textit{two individual documents} (or two sets of documents). We refer to the task as Comparative Document Analysis (CDA): Given a pair of documents from a document collection, the task is to (1) extract from each document salient phrases and phrase pairs which cover its major content; (2) discover the commonalities between the document pair by selecting salient phrases which are representative to \textit{both} of them; and (3) find the distinctions for each document by selecting salient phrases that are \textit{exclusively} relevant to the document.
CDA can benefit a variety of applications including related item recommendation and document retrieval. For example, as shown in Fig.~\ref{figure:motivated_example}, 
a citation recommendation system can show users the common and distinct concepts produced by CDA to help them understand the connections and differences between a query paper~\cite{jeh2003scaling} and a recommended paper~\cite{haveliwala2003topic}. In a similar way, CDA can reduce the efforts on patentbility searching~\cite{zhang2015patentcom}.
To give another example, in product recommendation scenario, CDA can address a user's doubts like ``why Amazon recommends \textit{Canon 6D camera} to me after I viewed \textit{Canon 5D3 camera}" by showing common aspects of the two cameras (\eg, ``\textit{DSLR camera}", ``\textit{full-frame sensor}") and the distinct aspects about \textit{Canon 6D} (\eg, ``\textit{WiFi function}"). 
By analyzing the common/distinct aspects and user's responses, a relevance feedback system can have a better understanding of the user's purchase intent and helps recommend other cameras with similar functions.

However, existing comparative summarization techniques encounter several unique challenges when solving CDA.

\begin{figure}
\centering
\vspace{-0.4cm}
\includegraphics[width = 41 mm]{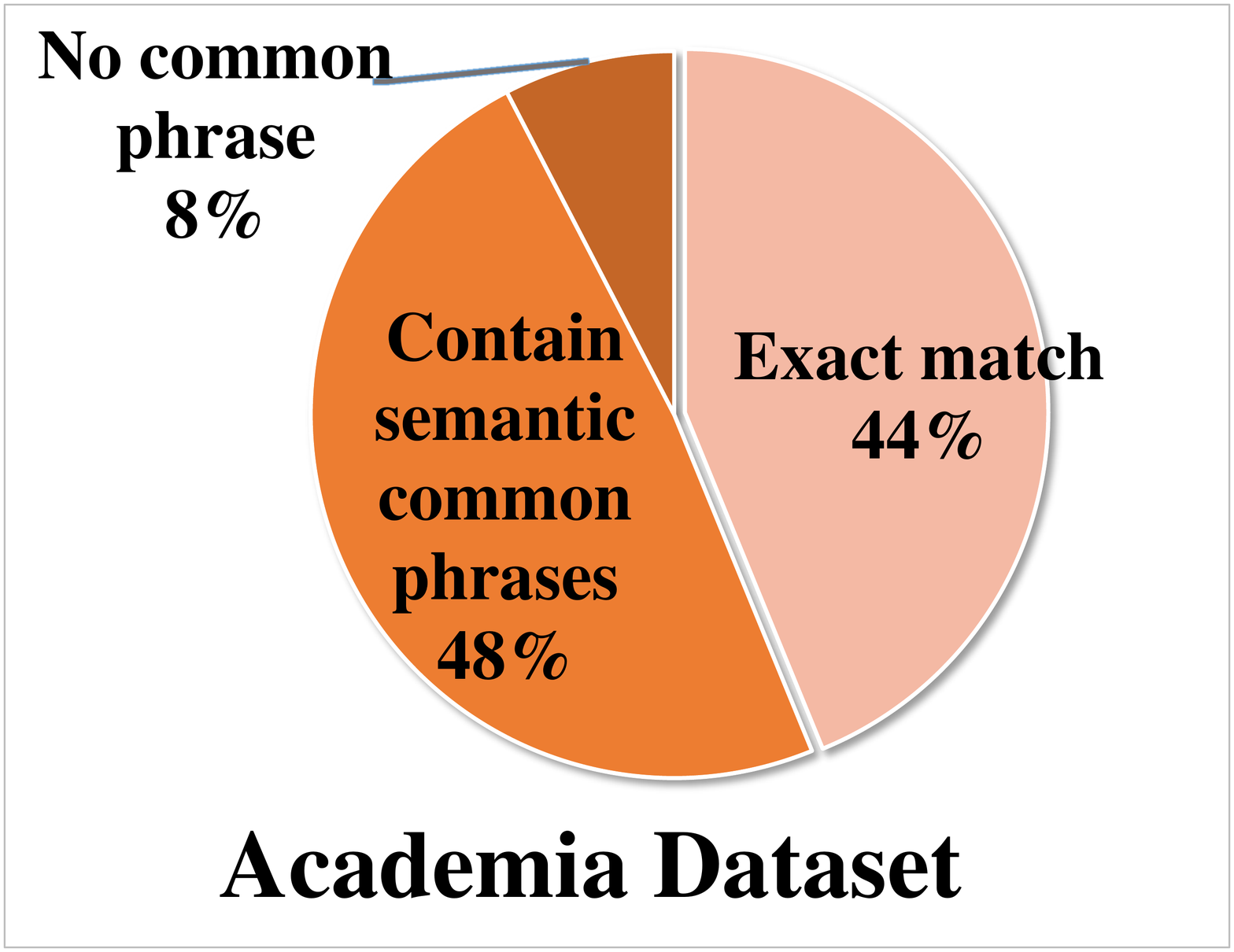}
\includegraphics[width = 41 mm]{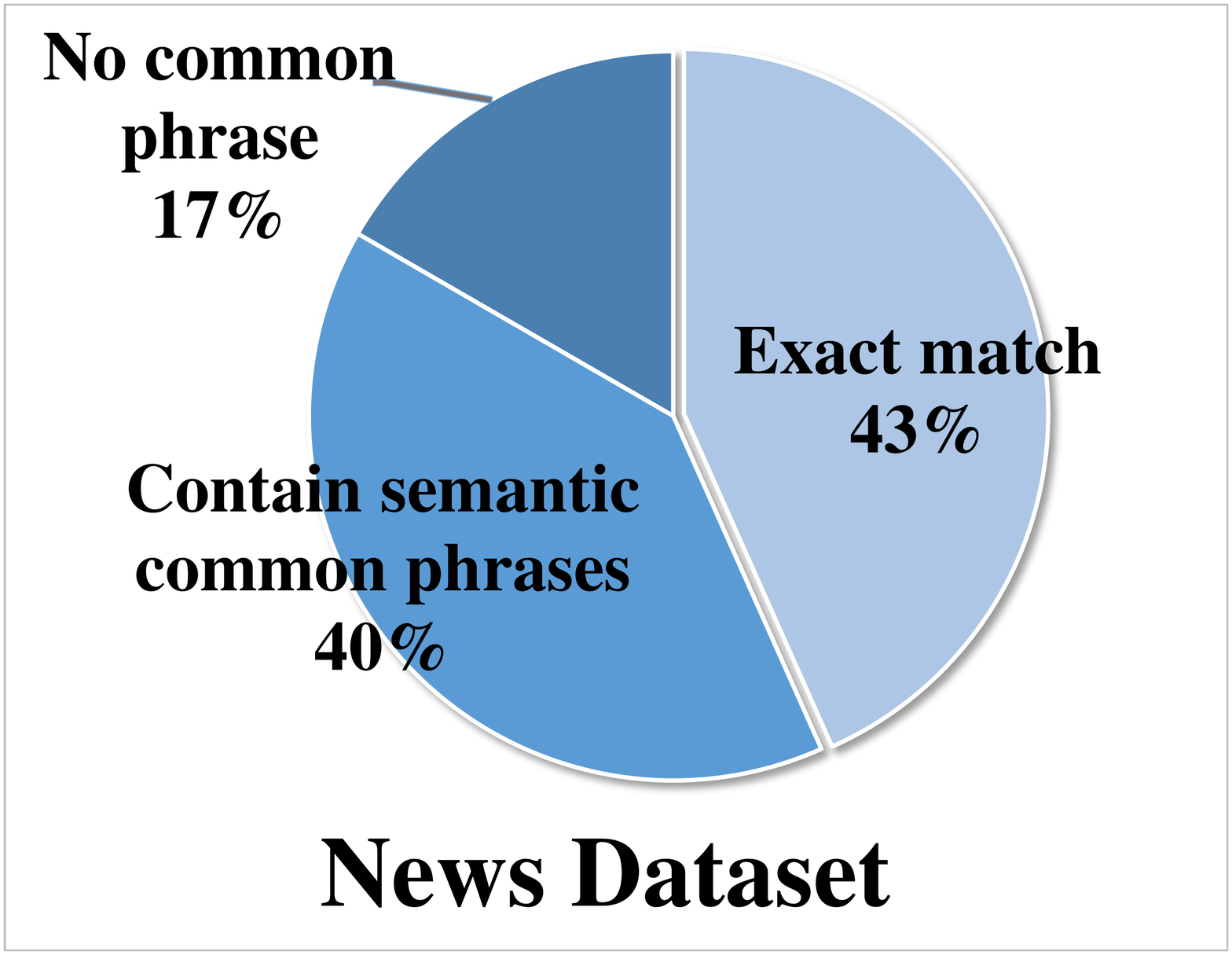}
\vspace{-0.2cm}
\caption{Study on semantic commonality: we show \%document pairs that contain \textit{semantic common phrases} (\ie, phrase which does not occur in \textit{both} document but represents the commonality) on the two evaluation sets. $62.37\%$ of the gold standard phrases are semantic common phrases.}
\label{figure:case_study_commonality}
\vspace{-0.2cm}
\end{figure}

\noindent$\bullet$ \textbf{Semantic Commonality:}
Commonalities between two documents can intuitively be bridged through phrases that do not \textit{explicitly} occur in both documents, \eg, 
``\textit{web graph\\@@web pages}" in Fig.~\ref{figure:sentence_document_graph}. Previous methods consider only content units (\eg, words) explicitly occurring in both documents as indication of commonalities but ignore such semantic commonality (see our study in Fig.~\ref{figure:case_study_commonality}).

\noindent$\bullet$ \textbf{Pairwise Distinction:}
Distinct phrases should be extracted based on the pair of compared documents \textit{jointly}---it should be \textit{exclusively} relevant to its own document (see Fig.~\ref{figure:proposed_measures}). Current methods select discriminative content units for each document set \textit{independently}. The results so generated thus may not distinguish the two documents effectively.

\noindent$\bullet$ \textbf{Data Sparsity:}
Most existing work relies on word overlap or sentence repetition between the items in comparison (\ie, document groups) to discover their commonalities and each item's distinctions. However, such sources of evidences may be absent when comparing two individual documents\footnote{\begin{small}One may argue that we can expand each individual document into a \textit{pseudo document collection} by retrieving its topic-related (item-related) documents. However, such a solution is expensive.\end{small}}. 

We address these challenges with several intuitive ideas.
First, to discover semantic commonalities between two documents, we consider phrases which are semantically relevant to \textit{both} documents as \textit{semantic common phrases} even they do not occur in both documents (\ie, (b) in Fig.~\ref{figure:proposed_measures}). 
Second, to select \textit{pairwise distinct phrases}, we use a novel measure that favors phrases relevant to one document but irrelevant to the other (\ie, (e) in Fig.~\ref{figure:proposed_measures}).
Third, to resolve data sparsity, we exploit phrase-document co-occurrence statistics in the entire corpus to go beyond the simple inter-document content overlap to model their semantic relevance. 

To systematically integrate these ideas, we propose a novel graph-based framework called \textbf{PhraseCom} (Fig.~\ref{figure:framework_overview}) to unify the formalization of optimization problems on selecting common phrases and distinct phrases. It first segments the corpus to extract candidate phrases where salient phrases for each document are selected based on phrase interestingness and diversity (Sec.~\ref{subsec:candidate_generation}). 
We then model the semantic relevance between phrases and documents using graph-based propagation over the co-occurrence graphs (Fig.~\ref{figure:sentence_document_graph}) to measure phrase commonality and distinction (Sec.~\ref{subsec:model}). We formulate two joint optimization problems using the proposed measures to select sets of common phrases and distinct phrases for a document pair, and present an alternative minimization algorithm to efficiently solve them (Sec.~\ref{subsec:optimization}-\ref{subsec:algorithm}). The algorithm tries to integrate the optimizing of the proposed commonality or distinction measure with learning of the semantic relevance scores, and can be flexibly extended to compare two sets of document (Sec.~\ref{subsec:algorithm}). 
The major contributions of this paper are summarized as follows.
\vspace{-0.0cm}
\begin{enumerate}[leftmargin=12pt]\itemsep+0.0cm
\item We define and study a novel comparative text mining task, comparative document analysis, which uses sets of phrases to jointly represent the commonality and distinctions between a pair (two sets) of documents.
\item We propose a graph-based framework, PhraseCom, to model and measure the semantic commonality and pairwise distinction for phrases in the documents. 
\item We formulate joint optimization problems to integrate the maximization of phrase commonality or distinction with the learning of phrase-document semantic relevance, and develop an efficient algorithm for solving them.
\item Experiments on datasets from different domains---news and academic papers---demonstrate that the proposed method achieves significant improvement over the state-of-the-art (\eg, 56\% enhancement in F1 on the Academia dataset over the best competitor from existing work).
\end{enumerate}

\vspace{-0.1cm}
\section{Problem Definition}
\label{subsec:problem}
\vspace{-0.0cm}

Literally, a comparison identifies the commonalities or differences among two or more objects. It consists of three components~\cite{wang2012comparative}: the \textbf{compared objects}, the \textbf{scale} (\ie, on what aspects the objects are measured and compared), and the \textbf{result}. 
In our task, compared objects are two given documents; scale is multi-word phrase or phrase pair; and the result is represented by three sets of phrases, \ie, a set of common phrases and two sets of distinct phrases for the two documents, respectively. 
The input to our proposed comparative document analysis framework is a document collection \begin{small}$\D=\{d_1,d_2,\ldots,d_n\}$\end{small}, a set of document pairs for comparison \begin{small}$\U=\{(d,\d')\}_{d,\d'\in\D}$\end{small}, and a set of positive example phrases \begin{small}$\Ppos=\{p^+_i,\dots,p^+_{n^+}\}$\end{small} extracted from knowledge bases for candidate phrase generation. In this work, we generate \begin{small}$\Ppos$\end{small} by using $\sim$500 Wikipedia article titles (see Sec.~\ref{subsec:candidate_generation} for details). 

A \textit{phrase}, $p$, is a single-word or multi-word sequence in the text document which represents a cohesive content unit (\eg, a noun phrase like ``\textit{automatic text summarizatio}n", or a verb phrase like ``\textit{earthquake struck}"). We further consider combining two phrases $p_a$ and $p_b$ into a \textit{phrase pair}, \ie, \begin{small}$p_a\oplus p_b$\end{small}, if they tend to co-occur with each other frequently in the document\footnote{\begin{small}Without explicitly mentioning, we refer both phrase and phrase pair as phrase in the rest of the paper.\end{small}} (\eg, ``\textit{web graph@@web pages}" in Fig.~\ref{figure:motivated_example}). 
Let \begin{small}$\P=\{p_1,\ldots,p_m\}$\end{small} denote $m$ unique phrases extracted from the corpus \begin{small}$\D$\end{small}.
For a pair of documents \begin{small}$(d, \d')\in\U$\end{small} in comparison, we denote the two sets of salient phrases extracted from them as \begin{small}$\S$\end{small} and \begin{small}$\Sp$\end{small}, respectively. We use a binary  vector \begin{small}$\mathbf{y}^c\in\{0,1\}^{m}$\end{small} to indicate whether salient phrases from the two documents (\ie, \begin{small}$\S\cup\Sp$\end{small}) are selected to form the set of common phrases \begin{small}$\C\subseteq \S\cup\Sp$\end{small}. Another two binary vectors, \begin{small}$\mathbf{y},\mathbf{y}^\prime\in\{0,1\}^{m}$\end{small}, are used to indicate whether salient phrases from \begin{small}$\S$\end{small} and \begin{small}$\Sp$\end{small} are selected as the set of distinct phrases, \ie, \begin{small}$\Q\subseteq\S$\end{small} and \begin{small}$\Qp\subseteq\Sp$\end{small}, for documents $d$ and $\d'$, respectively.
Obviously, \begin{small}$\C\cap\Q = \C\cap\Qp=\varnothing$\end{small}.
By estimating \begin{small}$\{\mathbf{y}^c, \mathbf{y}, \mathbf{y}^\prime\}$\end{small}, one can generate the three sets of phrases for the task, \ie, \begin{small}$\C = \{p~|~p\in\S\cup\Sp,~y^c_{p}=1\}$\end{small}, \begin{small}$\Q = \{p~|~p\in\S,~y_{p}=1\}$\end{small}, and \begin{small}$\Qp = \{p~|~p\in\Sp,~y^\prime_{p}=1\}$\end{small}.
Formally, we define the problem of \textbf{comparative document analysis (CDA)} as follows.
\vspace{-0.15cm}
\begin{definition}[Problem Definition]
Given a document collection \begin{small}$\D$\end{small}, a set of document pairs \begin{small}$\U$\end{small} and a set of positive example phrases \begin{small}$\Ppos$\end{small}, the task CDA aims to: (1) extract salient phrases \begin{small}$\S$\end{small} for each document \begin{small}$d\in\D$\end{small}; and (2) for each pair of comparing documents \begin{small}$(d, \d')\in\U$\end{small}, estimate the indicator vectors for the common and distinct phrases $\{\mathbf{y}^c, \mathbf{y}, \mathbf{y}^\prime\}$ to predict the comparison result sets $\{\C, \Q, \Qp\}$.
\end{definition}
In our study, we assume the given document pairs in $\U$ should be comparable, \ie, they share some common aspects or belong to the same concept category. For example, they can be two computer science papers or two news articles on similar topics. Such document pairs can come from document retrieval and item recommendation results. It is not the focus of this paper to generate such document pairs.

\section{Comparative Document Analysis}
\label{sec:method}
Our overall framework (shown in Fig.~\ref{figure:framework_overview}) is as follows:
\begin{enumerate}
\vspace{-0.15cm}
\item Perform distantly-supervised phrase segmentation on $\D$ and generate salient concepts $\S$ for $d\in\D$ by optimizing both phrase interestingness and diversity. (Sec.~\ref{subsec:candidate_generation}).

\vspace{-0.15cm}
\item Construct a phrase-document co-occurrence graph to help model semantic relevance as well as to assist measuring phrase commonality and distinction (Sec.~\ref{subsec:model}).

\vspace{-0.15cm}
\item Estimate indicator vectors $\{\mathbf{y}^c,\mathbf{y},\mathbf{y}^\prime\}$ for the three sets $\{\C, \Q, \Qp\}$ by solving the two proposed optimization problems with the efficient algorithms. (Sec.~\ref{subsec:optimization}-\ref{subsec:algorithm}).  
\end{enumerate}
\noindent
Each step will be elaborated in the following sections.

\begin{figure}[t]
\centering
\vspace{-0.4cm}
\includegraphics[width = 90mm]{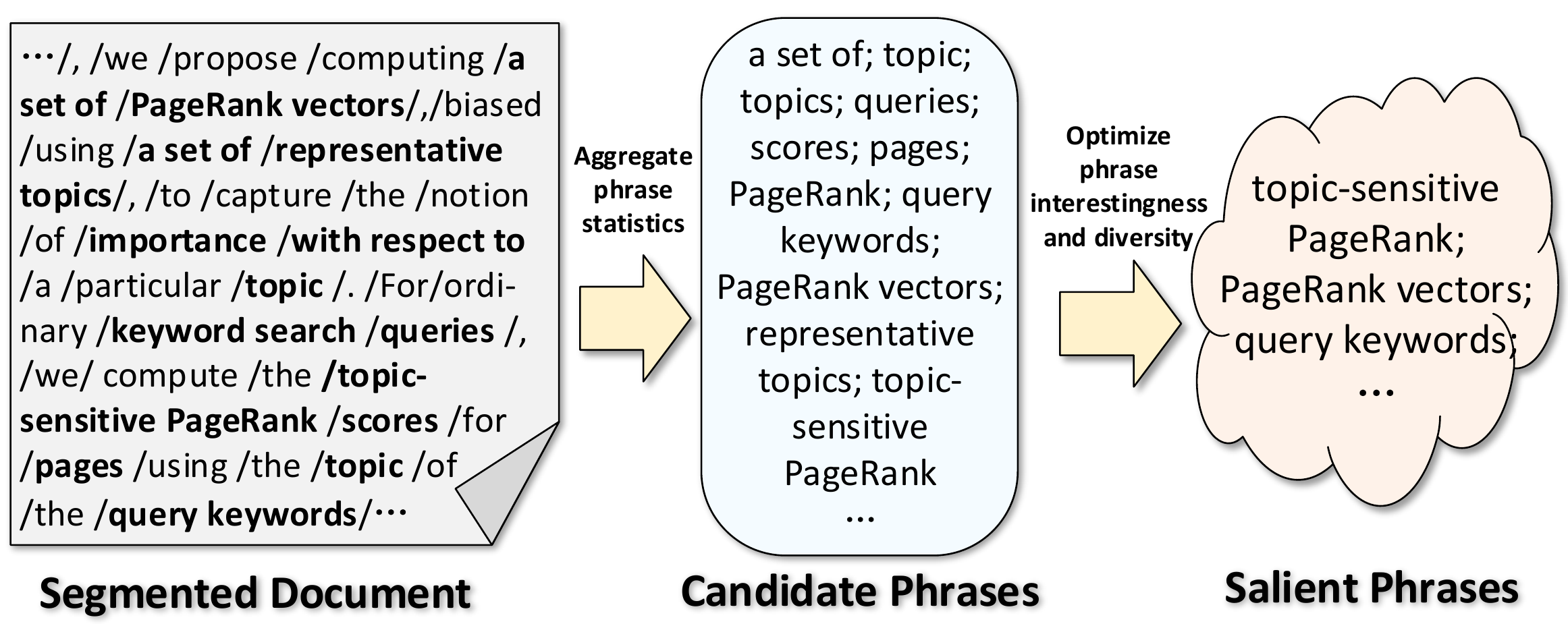}
\vspace{-0.4cm}
\caption{Output for \cite{haveliwala2003topic} in salient phrase generation.}
\label{figure:salient_phrase_generation}
\vspace{-0.2cm}
\end{figure}

\subsection{Salient Phrase Generation}
\label{subsec:candidate_generation}
To ensure the generation of cohesive, informative, salient phrases for each document, we introduce a scalable, data-driven approach by incorporating both local syntactic signals and corpus-level statistics. Our method first uses a distantly-supervised phrase segmentation algorithm~\cite{liu2015mining} to partition the text into non-overlapping segments for candidate phrase generation. Then it adopts both \textit{interestingness} and \textit{diversity} measures in a joint optimization framework~\cite{he2012gender} to guide the filtering of low importance phrases and removing phrases with redundant meaning.

Given a word sequence, the result of phrase segmentation is a sequence of phrases each representing a cohesive content unit (see Fig.~\ref{figure:salient_phrase_generation}).
First, we mine frequent contiguous patterns \begin{small}$\{v\}$\end{small} up to a fixed length. A random forest classifier is then used to estimate the phrase quality \begin{small}$Q(v)$\end{small} based on multiple features and the distant supervision \begin{small}$\Ppos$\end{small} from Wikiepdia.
With the estimated \begin{small}$Q(v)$\end{small}, we derive the optimal segmentation of each sentence using Viterbi Training~\cite{liu2015mining}. The above two steps can further iterate between each other to improve the results~\cite{liu2015mining}.
To enhance readability, we further combine phrases with good concordance in the document (\ie, \begin{small}$p_a$\end{small} and \begin{small}$p_b$\end{small} co-occur more than 3 times in a window of 10 words in \begin{small}$d$\end{small}) into phrase pairs \begin{small}$p_a\oplus p_b$\end{small}.

The results from phrase segmentation represents each document as a \textit{bag of phrases}, but the majority are not representative for the document. To select a set of salient phrases for a document, we consider two different aspects to measure the phrase salience, \ie, phrase \textit{interestingness} and phrase \textit{diversity}. The intuition behind phrase interestingness is simple~\cite{bedathur2010interesting}: a phrase is more interesting to the document if it appears frequently in the current document while relatively infrequently in the entire collection. Let \begin{small}$\P_d$\end{small} denote the set of phrases from the segmented document \begin{small}$d$\end{small} (including phrase pairs), \begin{small}$n(p,d)$\end{small} denote the frequency of \begin{small}$p$\end{small} in \begin{small}$d$\end{small}, and \begin{small}$n(p, \D)$\end{small} denote the document frequency of \begin{small}$p$\end{small} in \begin{small}$\D$\end{small}. The interestingness measure \begin{small}$r(\cdot)$\end{small} of \begin{small}$p$\end{small} in \begin{small}$d\in\D$\end{small} is defined as follows~\cite{bedathur2010interesting}.
\begin{align}
\label{eq:interestingness}
r_\D(p,d) = \Big(0.5+\frac{0.5 \times n(p, d)}{\max_{t\in \P_d}n(t,d)}\Big)^2 \cdot \log\Big(\frac{|\D|}{n(p,\D)}\Big),
\end{align}
which is the product of the square of normalized term frequency and the inverse document frequency.
In particular, interestingness score of phrase pair \begin{small}$p_a\oplus p_b$\end{small} is computed using an intra-document point-wise mutual information and is discount by its document frequency as follows.
\begin{align}
\label{eq:interestingness}
r_\D(p_a\oplus p_b,d) = \frac{\frac{n(p_a\oplus p_b,d)}{|\P_{d}|}}{\frac{n(p_a,d)}{|\P_{d}|}\frac{n(p_b,d)}{|\P_{d}|}} \cdot \log\Big(\frac{|\D|}{n(p_a\oplus p_b,\D)}\Big).
\end{align}
Interestingness scores for both phrases and phrase pairs from a document are normalized into \begin{small}$(0,1)$\end{small} for comparison.

To impose good diversity on the set of selected phrases, we enforce them to be  different from each other. We adopt edit distance-based \textit{Levenshtein similarity} to measure the string similarity \begin{small}$M(p_a, p_b)$\end{small} between two phrases \begin{small}$p_a$\end{small} and \begin{small}$p_b$\end{small}. 
One can also apply semantic similarity measures like distributional similarity~\cite{harris1954distributional}.
To select a subset \begin{small}$\S\subset\P_d$\end{small} of \begin{small}$K$\end{small} salient phrases for document \begin{small}$d$\end{small}, we solve an optimization problem~\cite{he2012gender} which maximizes interestingness and diversity \textit{jointly} as follows.
\begin{align}
\label{eq:phrase_optimization}
\argmaxl_{|\S|=K}~\H(\S) = \mu\sum_{p_a\in\P_d}q_a r_a - \sum_{p_a,p_b\in\P_d}r_aM_{ab}r_b.
\end{align}
Here, \begin{small}$r_a=r_\D(p_a, d)$\end{small} is the interestingness score, \begin{small}$M_{ab}=M(p_a, p_b)$\end{small} is phrase similarity score, and $q_a=\mathbf{M}_a \mathbf{r}$ is the importance score of \begin{small}$p_a$\end{small}. A near-optimal solution of Eq.~\eqref{eq:phrase_optimization} can be obtained by an efficient algorithm with time cost \begin{small}$\O(|\P_{d}|^2+|\P_{d}|K)$\end{small}~\cite{he2012gender}. Fig.~\ref{figure:salient_phrase_generation} provides an illustration of the salient phrase generation process with examples. 

\begin{figure}
\centering
\begin{small}
\vspace{-0.4cm}
\includegraphics[width = 72 mm]{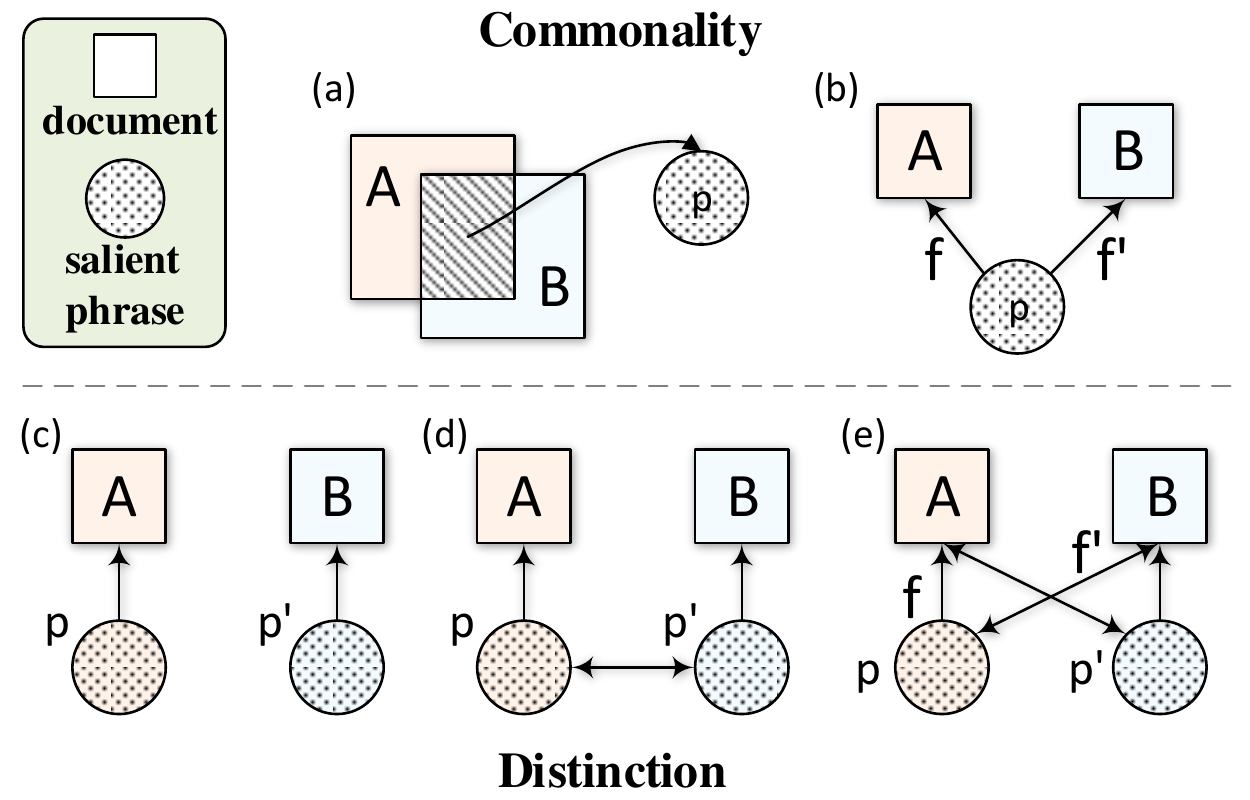}
\vspace{-0.2cm}
\caption{An illustration of the proposed measures. (a)~Intersection model; (b)~Semantic commonality model (ours): phrase is relevant to both documents; (c)~Independent distinction model; (d)~Joint distinction model: similar phrases are excluded; (e)~Pairwise distinction model (ours): phrase is \textit{relevant} to one document while \textit{irrelevant} to the other.}
\label{figure:proposed_measures}
\end{small}
\vspace{-0.2cm}
\end{figure}

\subsection{Commonality and Distinction Measures}
\label{subsec:model}
\begin{figure}
\centering
\begin{small}
\vspace{-0.2cm}
\includegraphics[width = 71 mm]{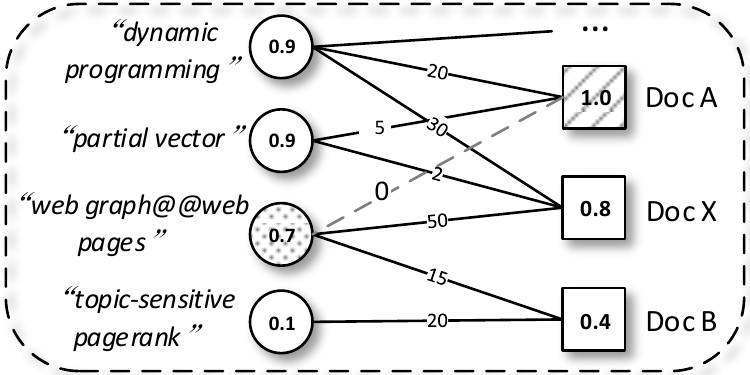}
\vspace{-0.0cm}
\caption{An illustration of relevance scores to Doc A derived from the constructed bipartite graph. Our method can infer the relevance between phrase ``\textit{web graph@@web pages}" and Doc A (\ie, 0.7) even it does not occur in Doc A.}
\label{figure:sentence_document_graph}
\end{small}
\vspace{-0.2cm}
\end{figure}

The salient phrase generation step provides a set of quality salient phrases \begin{small}$\S$\end{small} for each document \begin{small}$d\in\D$\end{small}. A straightforward solution to generate \begin{small}$\{\C,\Q,\Qp\}$\end{small} for CDA is to use phrases which are salient phrases for both documents to represent commonalities between the two documents (\ie, \begin{small}$\C = \S\cap\Sp$\end{small}) and to select the most salient phrases from the remaining ones to highlight the distinction (\ie, \begin{small}$\Q \subset \S\setminus\C,~\Qp \subset \Sp\setminus\C$\end{small})~\cite{mani1997summarizing}. However, such a solution have several major limitations. 
First, the common phrases so obtained ignore semantic common phrases, \ie, phrases which do not occur in the salient phrase sets of both documents but can also represent the commonalities (\eg, ``\textit{web graph@@web pages}" in Fig.~\ref{figure:sentence_document_graph}). Ignoring such semantic common phrases leads to low recall (see Fig.~\ref{figure:case_study_commonality}). 
Second, the simple selection of distinct phrases by excluding the common phrases only considers its relevance to the current document while ignoring its relationship to the other document. On one hand, this may not guarantee the pairwise distinction property---a salient phrase in one document may be also relevant to the other document. On the other hand, this often misses good distinct phrases which, although moderately salient in the current document, are completely irrelevant to the other document. 
Third, the distinct phrases so generated may include overly specific phrases which are salient in the current document but irrelevant to the other one (\eg, ``\textit{partial vectors}" in Fig.~\ref{figure:sentence_document_graph}). Recent work~\cite{zhang2015patentcom} considers pairwise distinction property by casting the problem as feature selection but it appears to produce many overly specific phrases. 

To systematically resolve these issues, we derive semantic relevance between phrases and documents based on their corpus-level co-occurrences statistics (Sec.~\ref{subsec:optimization}), and formalize novel objectives to measure semantic commonality and pairwise distinction for phrases.
The intuitive ideas are simple: (1) a good phrase to represent the commonality between two documents should be \textit{semantically} relevant to \textit{both} documents; (2) a good phrase to highlight the distinction of a document should be \textit{relevant} to this document but \textit{irrelevant} to the other document; and (3) a good phrase should be well accepted in the corpus (\ie, with reasonable popularity).

Specifically, let function \begin{small}$f(p,d):\P \times \D\mapsto \R_0^+$\end{small} denote the non-negative relevance score between \begin{small}$p\in\P$\end{small} and \begin{small}$d\in\D$\end{small}, which will be elaborated in Sec.~\ref{subsec:optimization}. 
We define the commonality score function \begin{small}$\Phi(p, d, \d'):\P \times \D\times \D\mapsto \R_0^+$\end{small} to measure how well phrase \begin{small}$p$\end{small} can represent the commonality between the document pair \begin{small}$(d,\d')$\end{small}. The following hypothesis guides our modeling of commonality score.
\begin{hypothesis}[Phrase Commonality]
\label{hypo:phrase_commonality}
Given \begin{small}$(d,\d')$\end{small}, phrase \begin{small}$p$\end{small} tends to have high commonality score \begin{small}$\Phi(p, d, \d')$\end{small} if and only if the relevance scores between the phrase and both documents, \ie, \begin{small}$f(p,d)$\end{small} and \begin{small}$f(p,\d')$\end{small}, are high.
\end{hypothesis}
In Fig.~\ref{figure:sentence_document_graph}, for example, ``\textit{web graph@@web pages}" has high commonality score since it has high relevance score to Doc B (it occurs frequently in Doc B) and high relevance score to Doc A (it occurs frequently in Doc X and Doc X contains several phrase that are relevant to Doc A). Formally, we define the \textbf{commonality score function} as follows.
\begin{align}
\label{eq:commonality_functuon}
\Phi(p,d,\d')=\ln\Big(1 + f(p,d)\cdot f(p,\d')\Big).
\end{align}
Similar to the product-of-experts model~\cite{hinton2002training}, it models the commonality score as the product of the two relevance scores, likes an ``and" operation. An alternative definition for \begin{small}$\Phi(\cdot)$\end{small} is the summation of two relevance scores, \ie, \begin{small}$\Phi(p,d,\d')= f(p,d)+ f(p,\d')$\end{small}. However, as an ``or" operation, the score so modeled may be dominated by the larger relevance score among the two. For instance, the case \begin{small}$f(p,d)=f(p,\d')=0.5$\end{small} and the case \begin{small}$f(p,d)=0.1, f(p,\d')=0.9$\end{small} share the same commonality score, but the former represents better commonality. We compare these two models in our experiments.

A good phrase for highlighting the distinction of \begin{small}$d$\end{small} between \begin{small}$(d,\d')$\end{small} should not only help distinguish \begin{small}$d$\end{small} from \begin{small}$\d'$\end{small} but also have good readability, \ie, not overly specific.
For example, in Fig.~\ref{figure:sentence_document_graph}, ``\textit{dynamic programming}" serves as a good distinct phrase for Doc A---it has strong association with Doc A and weak association with Doc B; meanwhile, it is fairly popular in the corpus. On the other hand, ``\textit{partial vector}" has similar association pattern with Doc A and Doc B but it is rarely mentioned in the corpus, \ie, overly specific. We use a distinction score function \begin{small}$\Pi(p, d,|\d'):\P \times \D\times \D\mapsto \R$\end{small} to measure how well phrase \begin{small}$p$\end{small} can highlight the distinction of \begin{small}$d$\end{small} from \begin{small}$\d'$\end{small}, based on the following hypothesis.
\begin{hypothesis}[Phrase Distinction]
\label{hypo:phrase_distinction}
Given \begin{small}$(d,\d')$\end{small}, \\phrase \begin{small}$p$\end{small} tends to have high distinction score \begin{small}$\Pi(p, d|\d')$\end{small} if it has relatively higher relevance score \begin{small}$f(p,d)$\end{small} to document \begin{small}$d$\end{small} compared with its relevance score \begin{small}$f(p,\d')$\end{small} to document \begin{small}$\d'$\end{small}.
\end{hypothesis}

Specifically, we define the \textbf{distinction score function} \begin{small}$\Pi(p, d|\d')$\end{small} based on the division of relevance to \begin{small}$d$\end{small} by the relevance to \begin{small}$\d'$\end{small}, which has the following form.
\begin{align}
\label{eq:distinction_functuon}
\Pi(p,d|\d')=\ln\Big(\frac{f(p,d)+\gamma}{f(p,\d')+\gamma}\Big).
\end{align}
Here, a smoothing parameter \begin{small}$\gamma=1$\end{small} is used to avoid selecting phrase \begin{small}$p$\end{small} with too small \begin{small}$f(p,d)$\end{small} or \begin{small}$f(p,\d')$\end{small}. In particular, the relevance score \begin{small}$f(p,d)$\end{small} incorporates the popularity of phrase \begin{small}$p$\end{small} in the collection (see Sec.~\ref{subsec:optimization}) and thus can filter overly specific phrases.
A phrase will receive high distinct score in two cases: (1) \begin{small}$p$\end{small} has high relevance score to \begin{small}$d$\end{small} and moderate or low relevance score to \begin{small}$\d'$\end{small}; and (2) \begin{small}$p$\end{small} has moderate relevance score to \begin{small}$d$\end{small} and low relevance score to \begin{small}$\d'$\end{small}. The second case helps include more phrases to represent the distinctions even they are moderately relevant to its own document. An alternative way to define the distinction score is by the difference between two relevance scores, \ie, \begin{small}$\Pi(p,d|\d')=f(p,d)-f(p,\d')$\end{small}. Such score functions prefer the first case than the second one, and thus will suffer from low recall. We compare with this alternative measure in the experiments.

\subsection{Comparative Selection Optimization}
\label{subsec:optimization}
With the proposed commonality and distinction measures, we now are concerned of the following two questions: (1) how to derive the relevance score \begin{small}$f(p,d)$\end{small} between phrase and document; and (2) how to select the sets of phrases \begin{small}$\{\C,\Q,\Qp\}$\end{small}. 

\smallskip
\noindent \textsf{\textbf{Graph-Based Semantic Relevance.}}
A simple idea to compute \begin{small}$f(p,d)$\end{small} is to use bag-of-words similarity measures such as cosine similarity. However, the score so generated may not capture the semantic relatedness between them (see StringFuzzy in Sec.~\ref{sec:experiments}). Our solution leverages graph-based semi-supervised learning~\cite{zhou2004ranking,zhu2005semi} to model the semantic relevance between phrases and documents, and further integrates the relevance score learning with phrase selection. It treats the target document \begin{small}$d\in\D$\end{small} as positive label and tries to rank the phrases \begin{small}$\P$\end{small} and the remaining documents \begin{small}$\D\setminus d$\end{small} based on the intrinsic manifold structure among them. 

By exploiting the aggregated co-occurrences between phrases and their supporting documents (\ie, documents where the phrase occurs) across the corpus \begin{small}$\D$\end{small}, we weight the importance of different phrases for a document, and use their connected edge as bridges to propagate the relevance scores between phrases and documents. The following hypothesis guides the modeling of relevance score \begin{small}$f(p,d)$\end{small}.

\begin{hypothesis}[Document-phrase Relevance]
\label{hypo:document_phrase_relevance}
If a phrase often appears in documents that are relevant to the target document, then the phrase tends to have high relevance score to the target document; If a document contains many phrases that are relevant to the target document, the document is likely relevant to the target document.
\end{hypothesis}
In Fig.~\ref{figure:sentence_document_graph}, for example, if we know ``\textit{dynamic programming}" and ``\textit{partial vector}" have high relevance scores regarding the target document Doc A, and find that the two phrases have strong association with Doc X, then Doc X is likely relevant to Doc A. This reinforces the relevance score propagation that ``\textit{web graph@@web page}" is also relevant to Doc A.

Specifically, we construct a bipartite graph \begin{small}$G$\end{small} to capture the co-occurrences between all the phrases \begin{small}$\P$\end{small} and documents \begin{small}$\D$\end{small}. A bi-adjacency matrix \begin{scriptsize}$\mathbf{W}\in\R_0^{+m\times n}$\end{scriptsize} is used to represent the edge weights for the links where \begin{small}$W_{ij}$\end{small} is the bag-of-words ranking score BM25~\cite{manning2008introduction} (\begin{small}$k_1=1.2,~b=0.75$\end{small}) between \begin{small}$p_i\in\P$\end{small} and \begin{small}$d_j\in\D$\end{small} if \begin{small}$p_i$\end{small} occurs in \begin{small}$d_j$\end{small}, and otherwise zero. We further normalize \begin{small}$\mathbf{W}$\end{small} by the popularity of the phrases and documents to reduce the impact of overly popular phrases~\cite{zhu2005semi}.
\vspace{-0.1cm}
\begin{align}
\label{eq:normalized_bipartite_graph}
\mathbf{S} = \mathbf{D}^{(\P)-\frac{1}{2}}\cdot\mathbf{W} \cdot \mathbf{D}^{(\D)-\frac{1}{2}},
\end{align}
where we define the diagonal matrices \begin{scriptsize}$\mathbf{D}^{(\P)}\in\R^{m\times m}$\end{scriptsize} as \begin{scriptsize}$D^{(\P)}_{ii}=\sum_{j=1}^n W_{ij}$\end{scriptsize} and \begin{scriptsize}$\mathbf{D}^{(\D)}\in\R^{n\times n}$\end{scriptsize} as \begin{scriptsize}$D^{(\D)}_{jj}=\sum_{i=1}^m W_{ij}$\end{scriptsize}.

Formally, we use function \begin{scriptsize}$g(d_a, d_b):\D\times \D\mapsto\R_0^{+}$\end{scriptsize} to denote the relevance score between two documents \begin{small}$d_a, d_b\in\D$\end{small}, and define vector \begin{scriptsize}$\mathbf{f}\in\R_0^{+^m}$\end{scriptsize} as \begin{small}$f_i = f(p_i,d)$\end{small} and vector \begin{scriptsize}$\mathbf{g}\in\R_0^{+^n}$\end{scriptsize}as \begin{small}$g_j = g(d_j,d)$\end{small}. Following Hypothesis~\ref{hypo:document_phrase_relevance}, we use target document \begin{small}$d$\end{small} as positive label and model the relevance score propagation between phrases and documents using a graph regularization term and a supervision term~\cite{zhou2004ranking}.
\begin{small}
\begin{align}
\label{eq:relevance_score_propagation}
\L_{d,\alpha}(\mathbf{f},\mathbf{g}) = \sum_{i=1}^m\sum_{j=1}^n W_{ij}\Big(\frac{f_i}{\sqrt{D_{ii}^{(\P)}}} - \frac{g_j}{\sqrt{D_{jj}^{(\D)}}}\Big)^{2}+\alpha\|\mathbf{g}-\mathbf{g}^0\|_2^2.
\end{align}
\end{small}
Here, we define the indicator vector \begin{small}$\mathbf{g}^0\in\{0,1\}^n$\end{small} to impose the positive label of \begin{small}$d$\end{small} in the second term, where \begin{small}$g^0_d=1$\end{small} and \begin{small}$0$\end{small} otherwise. A tuning parameter \begin{small}$\alpha>0$\end{small} is used to control the strength of supervision from \begin{small}$d$\end{small} on the score propagation.

\smallskip
\noindent \textsf{\textbf{The Joint Optimization Problems.}}
Relevance score \begin{small}$f(p,d)$\end{small} can be directly learned by minimizing \begin{small}$\L_d(\mathbf{f},\mathbf{g})$\end{small} but the score so computed is \textit{document independent}---it considers only information from \begin{small}$d$\end{small} while ignoring that from the other \begin{small}$\d'$\end{small} in the compared pair \begin{small}$(d,\d')$\end{small}. To address this problem, we propose two joint optimization problems for selecting common phrases and distinct phrases, respectively, by incorporating information from both documents. The intuition is simple: phrases which are likely common (distinct) phrases can reinforce the propagation between relevance scores by serving as extra positive labels (\ie, besides the original document \begin{small}$d$\end{small}). In our experiments, we compare with the independent optimization method (\ie, CDA-TwoStep).

Formally, to discover the common phrases \begin{small}$\C$\end{small} between a document pair \begin{small}$(d,\d')\in\U$\end{small}, we propose the \textbf{common phrase selection problem} which unifies two different tasks: (i) selection of \begin{scriptsize}$\C\subset \S\cup\Sp$\end{scriptsize} to maximize the overall commonality score; and (ii) minimization of the graph-based objectives to learn relevance scores.
Let \begin{scriptsize}$\mathbf{f}^\prime\in\R_0^{+^m}$\end{scriptsize} denote phrase-document relevance scores \begin{small}$f(p_i,\d')$\end{small} for \begin{small}$p_i\in\P$\end{small}, and \begin{scriptsize}$\mathbf{g}^\prime\in\R_0^{+^n}$\end{scriptsize} denote document-document relevance scores \begin{small}$g(d_j,\d')$\end{small} for \begin{small}$d_j\in\D$\end{small}, we formulate the problem as follows.
\begin{footnotesize}
\begin{align}
\label{eq:commonality_optimization}
& \min_{\mathbf{y}^c,\mathbf{f},\mathbf{f}^\prime,\mathbf{g},\mathbf{g}^\prime} \O_{\alpha,\lambda} 
 =-\lambda\sum_{i=1}^m y^c_i\cdot \Phi(p_i,d,\d')\\
& \hspace{2.4cm}+\frac{1}{2}\L_{d,\alpha}(\mathbf{f},\mathbf{g})+\frac{1}{2}\L_{\d',\alpha}(\mathbf{f}^\prime,\mathbf{g}^\prime) \nonumber\\
& \st \quad y^c_i\cdot\Phi(p_i,d,\d')\geq y^c_i\sum_{p_j\in\S}\Phi(p_j,d,\d')/|\S|,~~\forall p_i\in\P; \nonumber\\
& ~\quad \quad  y^c_i\cdot\Phi(p_i,d,\d')\geq y^c_i\sum_{p_j\in\Sp}\Phi(p_j,d,\d')/|\Sp|,~~\forall p_i\in\P; \nonumber\\
& ~\quad \quad\|\mathbf{y}^c - \mathbf{y}^{\S\cup\Sp}\|_2 \leq \|\mathbf{y}^{\S\cup\Sp}\|_2. \nonumber
\end{align}
\end{footnotesize}
The first term in objective \begin{small}$\O$\end{small} aggregates the commonality scores for the selected phrases, and the tuning parameter \begin{small}$\lambda>0$\end{small} controls the trade-off between it and the second and third terms which model the relevance score propagation on graph. We add the first and second constraints to automatically decide the size of \begin{small}$\C$\end{small}, by enforcing that the selected phrases should have higher commonality score than the average commonality scores computed over salient phrases \begin{small}$\S$\end{small} and \begin{small}$\Sp$\end{small}, respectively. The last constraint enforces \begin{small}$\C\subset\S\cup\Sp$\end{small}, where \begin{scriptsize}$y^{\S\cup\Sp}_i=1$\end{scriptsize} if \begin{scriptsize}$p_i\in\S\cup\Sp$\end{scriptsize}, and zero otherwise.

To jointly select the distinct phrases \begin{scriptsize}$\{\Q,\Qp\}$\end{scriptsize} for pair \begin{small}$(d,\d')\in\U$\end{small} , we propose the \textbf{distinct phrase selection problem}. It aims to: (i) select phrases \begin{scriptsize}$\Q\subset\S$\end{scriptsize} and \begin{scriptsize}$\Qp\subset\Sp$\end{scriptsize} to maximize the overall distinction scores; and (ii) derive relevance scores by minimizing the graph-based objectives. Specifically, we formulate the problem as follows.
\begin{footnotesize}
\begin{align}
\label{eq:distinction_optimization}
& \min_{\mathbf{y},\mathbf{y}^\prime,\mathbf{f},\mathbf{f}^\prime,\mathbf{g},\mathbf{g}^\prime} \F_{\alpha,\lambda} 
 =-\lambda\sum_{i=1}^m \Big[y_i\cdot\Pi(p_i,d|\d') + y_i^\prime \cdot\Pi(p_i,\d'|d)\Big]\nonumber \\
& \hspace{2.6cm}+\frac{1}{2}\L_{d,\alpha}(\mathbf{f},\mathbf{g})+\frac{1}{2}\L_{\d',\alpha}(\mathbf{f}^\prime,\mathbf{g}^\prime) \\
& \st \quad y_i\cdot\Pi(p_i,d|\d')\geq y_i\sum_{p_j\in\S}\Pi(p_j,d|\d')/|\S|,~~p_i\in\P; \nonumber\\
& ~\quad \quad  y^\prime_i\cdot\Pi(p_i,\d'|d)\geq y^\prime_i\sum_{p_j\in\Sp}\Phi(p_j,\d'|d)/|\Sp|,~~p_i\in\P; \nonumber\\
& ~\quad \quad\|\mathbf{y} - \mathbf{y}^{\S}\|_2 \leq \|\mathbf{y}^{\S}\|_2,~
\|\mathbf{y}^\prime - \mathbf{y}^{\Sp}\|_2 \leq \|\mathbf{y}^{\Sp}\|_2 \nonumber\\
& ~\quad \quad\|\mathbf{y} - \mathbf{y}^c\|_2 = \|\mathbf{y}\|_2 + \|\mathbf{y}^c\|_2,~\|\mathbf{y}^\prime - \mathbf{y}^c\|_2 = \|\mathbf{y}^\prime\|_2 + \|\mathbf{y}^c\|_2. \nonumber
\end{align}
\end{footnotesize}
The first term represents the aggregated distinction score over the two sets of selected distinct phrases, \ie, \begin{small}$\Q$\end{small} and \begin{small}$\Qp$\end{small}, respectively. Similarly, the first two constraints help control the output size and the third constraint enforces \begin{small}$\Q\subset\S,~\Qp\subset\Sp$\end{small}. In particular, the last constraint excludes the selected common phrases \begin{small}$\C$\end{small} from the results \begin{small}$\Q,~\Qp$\end{small}.

\subsection{An Efficient Algorithm}
\label{subsec:algorithm}
The optimization problems in Eq.~\eqref{eq:commonality_optimization} and Eq.~\eqref{eq:distinction_optimization} are mix-integer programming and thus are NP-hard. We propose an approximate solution for each problem based on the alternative minimization framework~\cite{tseng2001convergence}: first estimate the relevance scores \begin{small}$\{\mathbf{f},\mathbf{f}^\prime,\mathbf{g},\mathbf{g}^\prime\}$\end{small} through minimizing \begin{small}$\O$\end{small} while fixing \begin{small}$\mathbf{y}^c$\end{small} (or \begin{small}$\{\mathbf{y},\mathbf{y}^\prime\}$\end{small}); then fix \begin{small}$\{\mathbf{f},\mathbf{f}^\prime,\mathbf{g},\mathbf{g}^\prime\}$\end{small} and optimize \begin{small}$\O$\end{small} with respect to \begin{small}$\mathbf{y}^c$\end{small} (or \begin{small}$\{\mathbf{y},\mathbf{y}^\prime\}$\end{small}) by imposing the constraints; and iterate between these two steps until reaching the convergence of the objective function \begin{small}$\O$\end{small}. 

For common phrase selection problem in Eq.~\eqref{eq:commonality_optimization}, to estimate the relevance score vectors \begin{small}$\{\mathbf{f},\mathbf{f}^\prime,\mathbf{g},\mathbf{g}^\prime\}$\end{small}, we take derivative on \begin{small}$\O$\end{small} with respect to each of the variables in\begin{small}$\{\mathbf{f},\mathbf{f}^\prime,\mathbf{g},\mathbf{g}^\prime\}$\end{small} while fixing \begin{small}$\mathbf{y}^c$\end{small} and other variables. By setting the derivatives to zero, we have the following update rules.
\begin{small}
\begin{align}
f_i &= \left\{
  \begin{array}{lll}
    \frac{\mathbf{S}_i\mathbf{g}}{2} - \frac{1}{2 f_i^\prime}+\sqrt{(\frac{\mathbf{S}_i\mathbf{g}}{2}+\frac{1}{2f^\prime_i})^2+\lambda y^c_i}, & \hbox{if $f_i^\prime > 0$;} \\ 
    \mathbf{S}_i \mathbf{g}, & \hbox{if $f_i^\prime = 0$.}
  \end{array}
\right. \label{eq:update_common_f}\\
g_j & = (\mathbf{S}_{\cdot j}^T\mathbf{f}+\alpha\cdot g_j^0)/(1+\alpha) \label{eq:update_common_g}\\
f^\prime_i & = \left\{
  \begin{array}{lll}
    \frac{\mathbf{S}_i\mathbf{g}^\prime}{2} - \frac{1}{2 f_i}+\sqrt{(\frac{\mathbf{S}_i\mathbf{g}^\prime}{2}+\frac{1}{2f_i})^2+\lambda y^c_i}, & \hbox{if $f_i > 0$;} \\
    \mathbf{S}_i \mathbf{g}^\prime, & \hbox{if $f_i = 0$.}
  \end{array}
\right. \label{eq:update_common_fp}\\
g_j^\prime & = (\mathbf{S}_{\cdot j}^T\mathbf{f}^\prime+\alpha\cdot g_j^{\prime 0})/(1+\alpha), \label{eq:update_common_gp}
\end{align}
\end{small}
Here \begin{small}$\mathbf{S}_i$\end{small} and \begin{small}$\mathbf{S}_{\cdot j}$\end{small} denote $i$-th row and $j$-th column of matrix \begin{small}$\mathbf{S}$\end{small}, respectively. With the updated relevance score vectors \begin{small}$\{\mathbf{f},\mathbf{f}^\prime,\mathbf{g},\mathbf{g}^\prime\}$\end{small}, we can compute the commonality score \begin{small}$\Phi(p_i,d,\d')$\end{small} for \begin{small}$p_i\in\P$\end{small} and estimate the common phrase indicator vector \begin{small}$\mathbf{y}^c$\end{small} as follows.
\begin{align}
\label{eq:update_common_yc}
y^c_i = \left\{
  \begin{array}{lll}
    1, & \hbox{if $p_i\in\P$ satisfies contraints in Eq.~\eqref{eq:commonality_optimization};} \\
    0, & \hbox{otherwise.}
  \end{array}
\right.
\end{align}

\begin{algorithm}[t]
\begin{small}
\DontPrintSemicolon
\KwIn{document pair $(d,\d')$, salient phrases $\{\S,\Sp\}$, bi-adjacency matrix $\mathbf{S}$, tuning parameters $\{\alpha,\lambda\}$}
\KwOut{common phrases $\C$, distinct phrases $\{\Q,\Qp\}$}
Initialize: $\{\mathbf{g}^0,\mathbf{g}^{\prime 0}\}$ by $\{d,\d'\}$; $\{\mathbf{y}^c,\mathbf{f},\mathbf{f}^\prime,\mathbf{g},\mathbf{g}^\prime\}$ as zero vectors\;
$\mathbf{g}^0\gets\mathbf{g}^0+0.1;~\mathbf{g}^{\prime 0}\gets\mathbf{g}^{\prime 0}+0.1$\;
\Repeat{the objective $\O$ in Eq.~\eqref{eq:commonality_optimization} converges}{
	\Repeat{Eq.~\eqref{eq:relevance_score_propagation} converges}{
      Update $\{\mathbf{f},\mathbf{g}\}$ using Eq.~\eqref{eq:update_common_f} and Eq.~\eqref{eq:update_common_g}\;
      Update $\{\mathbf{f}^\prime,\mathbf{g}^\prime\}$ using Eq.~\eqref{eq:update_common_fp} and Eq.~\eqref{eq:update_common_gp}\;
    	}
    	Update $\mathbf{y}^c$ following Eq.~\eqref{eq:update_common_yc} 
}
\Return{$\C$ based on the estimated $\mathbf{y}^c$}\;
Initialize: $\{\mathbf{y}, \mathbf{y}^\prime\}$ and $\{\mathbf{f},\mathbf{f}^\prime,\mathbf{g},\mathbf{g}^\prime\}$ as zero vectors\;
\Repeat{the objective $\F$ in Eq.~\eqref{eq:distinction_optimization} converges}{
	\Repeat{Eq.~\eqref{eq:relevance_score_propagation} converges}{
      Update $\{\mathbf{f},\mathbf{g}\}$ using Eq.~\eqref{eq:update_distinction_f} and Eq.~\eqref{eq:update_common_g}\;
      Update $\{\mathbf{f}^\prime,\mathbf{g}^\prime\}$ using Eq.~\eqref{eq:update_distinction_fp} and Eq.~\eqref{eq:update_common_gp}\;
    	}
    	Update $\{\mathbf{y},\mathbf{y}^\prime\}$ based on $\C$ and Eq.~\eqref{eq:update_distinct_y_yp}}
\Return{$\{\Q,\Qp\}$ based on the estimated $\{\mathbf{y},\mathbf{y}^\prime\}$}\;
\caption{{\sc PhraseCom}}
\label{algorithm:PhraseCom}
\end{small}
\end{algorithm}

To estimate the relevance score vectors for the distinct phrase selection problem in Eq~\eqref{eq:distinction_optimization}, we first optimize \begin{small}$\F$\end{small} with respect to \begin{small}$\{\mathbf{f},\mathbf{f}^\prime,\mathbf{g},\mathbf{g}^\prime\}$\end{small} while fixing \begin{small}$\{\mathbf{y}, \mathbf{y}^\prime\}$\end{small}. The update rule can be derived by taking derivative of \begin{small}$\F$\end{small} with respect to each variable in \begin{small}$\{\mathbf{f},\mathbf{f}^\prime,\mathbf{g},\mathbf{g}^\prime\}$\end{small} while fixing the other variables and setting the derivative to zero.
\begin{align}
f_i &= -\frac{\gamma-\mathbf{S}_i\mathbf{g}}{2} + \sqrt{\big(\frac{\gamma+\mathbf{S}_i\mathbf{g}}{2}\big)^2+\lambda(y_i-y^\prime_i)}  \label{eq:update_distinction_f}, \\
f^\prime_i & = -\frac{\gamma-\mathbf{S}_i\mathbf{g}^\prime}{2} + \sqrt{\big(\frac{\gamma+\mathbf{S}_i\mathbf{g}^\prime}{2}\big)^2+\lambda(y_i^\prime-y_i)}, \label{eq:update_distinction_fp}
\end{align}
In particular, update rules for \begin{small}$\{\mathbf{g}, \mathbf{g}^\prime\}$\end{small} are the same as Eq.~\eqref{eq:update_common_g} and Eq.~\eqref{eq:update_common_gp}. To ensure the existence of a solution and the non-negativity for \begin{small}$\{\mathbf{f}, \mathbf{f}^\prime\}$\end{small}, the tuning parameters need to satisfy the condition \begin{small}$\lambda \leq \min\big(\gamma^2/4,~\alpha\gamma\delta/(\alpha+1)\big)$\end{small}. Here, a small positive value \begin{small}$\delta=0.1$\end{small} is added to \begin{small}$\{\mathbf{g}^0, \mathbf{g}^{\prime 0}\}$\end{small} to help derive stable solution.
Similar to Eq.~\eqref{eq:update_common_yc}, the two indicator vectors \begin{small}$\{\mathbf{y}, \mathbf{y}^\prime\}$\end{small} can be estimated by first computing distinction scores \begin{small}$\Pi(p_i,d|\d')$\end{small} and \begin{small}$\Pi(p_i,\d'|d)$\end{small} based on the estimated \begin{small}$\{\mathbf{f},\mathbf{f}^\prime,\mathbf{g},\mathbf{g}^\prime\}$\end{small} and then enforcing constraints in Eq.~\eqref{eq:distinction_optimization}.
\begin{align}
\label{eq:update_distinct_y_yp}
y_i,~y^\prime_i = \left\{
  \begin{array}{lll}
    1, & \hbox{if $p_i\in\P$ satisfies the corresponding} \\
    & \hbox{contraints in Eq.~\eqref{eq:distinction_optimization};} \\
    0, & \hbox{otherwise.}
  \end{array}
\right. 
\end{align}

Algorithm~\ref{algorithm:PhraseCom} summarizes our algorithm \textbf{PhraseCom}. For convergence analysis, PhraseCom applies block coordinate descent on problems in Eq.~\eqref{eq:commonality_optimization} and Eq.~\eqref{eq:distinction_optimization}. The proof procedure in~\cite{tseng2001convergence} (not included for lack of space) can be adopted to prove convergence for PhraseCom (to the local minima).

\smallskip
\noindent \textsf{\textbf{Extension to compare two sets of documents.}} PhraseCom can be easily extended to perform the comparison between two sets of documents, \ie, \begin{small}$(\D_a, \D_b)$\end{small}. Let \begin{small}$\S_a=\cup_{d\in\D_a}\S_d$\end{small} and \begin{small}$\S_b=\cup_{d\in\D_b}\S_d$\end{small} denote the salient phrases extracted from the two sets of documents, respectively. Our method can directly replaces \begin{small}$\{\S,\Sp\}$\end{small} by \begin{small}$\{\S_a,\S_b\}$\end{small}, initializes the vectors \begin{small}$\{\mathbf{g}^0,\mathbf{g}^{\prime 0}\}$\end{small} based on \begin{small}$\{\D_a,\D_b\}$\end{small}, and derive the comparison results \begin{small}$\{\C,\Q_a,\Q_b\}$\end{small} following Algorithm~\ref{algorithm:PhraseCom}.

\smallskip
\noindent \textsf{\textbf{Computational Complexity Analysis.}}
Given a corpus \begin{small}$\D$\end{small} with \begin{small}$n$\end{small} documents and \begin{small}$N_\D$\end{small} words, the time cost for salient phrase generation is \begin{small}$\O(N_\D+n|\P_d|^2)$\end{small} where \begin{small}$|\P_d|$\end{small} denotes average number of candidate phrases in each document.
In phrase-document graph construction, the costs for computing \begin{small}$\mathbf{S}$\end{small} is \begin{small}$\O(N_\D+ n|\P_d|l_p)$\end{small} where \begin{small}$l_p$\end{small} is the average number of words in a phrase (\begin{small}$l_p\ll |\P_d|$\end{small}). The overall cost of data preparation is \begin{small}$\O(N_\D+ n|\P_d|^2)$\end{small}. In Algorithm~\ref{algorithm:PhraseCom}, it takes \begin{small}$\O(n+m)$\end{small} time to initialize all the variables.
Each iteration of the inner loop costs \begin{small}$\O(n|\P_d|+m)$\end{small} to update \begin{small}$\{\mathbf{f},\mathbf{f}^\prime\}$\end{small} and \begin{small}$\O(n|\P_d|)$\end{small} to update \begin{small}$\{\mathbf{g},\mathbf{g}^\prime\}$\end{small}. The cost for inner loop is \begin{small}$\O(n|\P_d|t_{i}+m t_i)$\end{small} supposing it stops after \begin{small}$t_i$\end{small} iterations (\begin{small}$t_{i}<50$\end{small} in our experiments). 
In the outer loop, update of \begin{small}$\{\mathbf{y}^c,\mathbf{y},\mathbf{y}^\prime\}$\end{small} costs \begin{small}$\O(m)$\end{small} time. Overall, suppose that the outer loop terminates in \begin{small}$t_o$\end{small} iterations, the computational complexity of PhraseCom is \begin{small}$\O\big(n|\P_d|t_{i}t_o+m t_i t_o\big)$\end{small} (\begin{small}$t_{o}<5$\end{small} in our experiments).


\section{Experiments}
\label{sec:experiments}

\subsection{Data Preparation}
\label{subsec:data_preparation}
\vspace{-0.0cm}
\nop{
\noindent
{\large \textbf{5.1. Data Preparation:}}
}
Our experiments use two real-world datasets\footnote{\begin{small}\url{http://dl.acm.org/};~\url{http://www.newsbank.com/}\end{small}}:
\begin{itemize}
\item \textbf{Academia:} We collected 205,484 full-text papers (574M tokens and 1.98M unique words) published in a variety of venues between 1990 and 2015 from ACM Digital Library.
\item \textbf{News:} constructed by crawling news articles published between Mar. 11 and April 11, 2011 with  keywords “Japan Tsunami” from NewsBank. This yields a collection of 67,809 articles (29M tokens and 247k unique words).
\end{itemize}

\renewcommand{\arraystretch}{1.2}
\begin{table}[t]
\vspace{-0.4cm}
\begin{small}
\begin{center}
\vspace{0.0cm}
\begin{tabularx}{\linewidth}{  x l l } \hline
\textbf{Data sets} & \textbf{Academia} & \textbf{News} \\ \hline
$\#$Documents & 205,484 & 67,809 \\ 
$\#$Candidate phrases & 611,538 & 224,468 \\ 
$\#$Salient phrases & 316,194 & 119,600 \\ 
$\#$Salient phrase pairs & 204,744 & 11,631 \\ 
$\#$Unique words & 1.98M & 246,606 \\ 
$\#$Links & 153.19M & 2.85M \\ 
$\#$Salient phrases per doc & 18.95 & 17.43 \\ 
$\#$Salient phrase pairs per doc & 2.42 & 1.31 \\ \hline
\end{tabularx}
\vspace{-0.1cm}
\caption{Statistics of the datasets.}
\label{table:data_stats}
\vspace{-0.1cm}
\end{center}
\end{small}
\end{table}

\begin{figure}[t]
\centering
\vspace{-0.0cm}
\includegraphics[width = 72 mm]{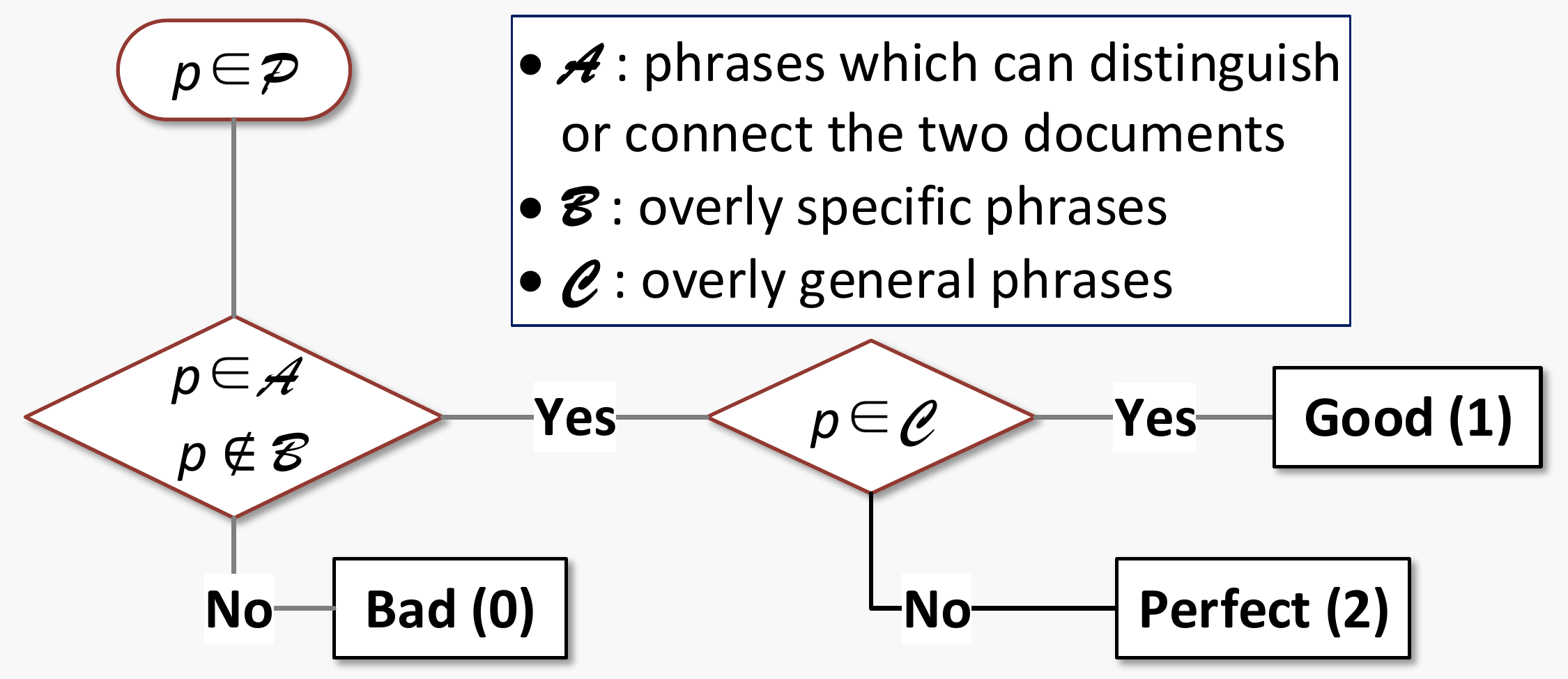}
\vspace{-0.1cm}
\caption{Annotation criterion.}
\label{figure:annotation_criterion}
\vspace{-0.2cm}
\end{figure}

\smallskip
\noindent
\textsf{\textbf{1. Salient Phrases.}}
We first performed lemmatization on the tokens to reduce variant forms of words (\eg, ate, eating) into their lemmatized form (\eg, eat).
In phrase segmentation, we set maximal pattern length to $5$, minimum support to $10$, and non-segmented ratio to $0.05$ in SegPhrase algorithm~\cite{liu2015mining}, to extract candidate phrases from the corpus.
We then used the GenDeR algorithm in \cite{he2012gender} to solve the salient phrase selection problem in Eq.~\eqref{eq:phrase_optimization}. We set weighting parameter \begin{small}$\mu=3$\end{small} and maximal number of salient phrase selected for each document \begin{small}$K=30$\end{small}. Member phrases in salient phrase pairs were removed from the salient phrase set.

\smallskip
\noindent
\textsf{\textbf{2. Bipartite Graphs.}}
We followed the introduction in Sec.~\ref{subsec:optimization} to construct the phrase-document bipartite graph for each dataset. To compute the BM25 score between a phrase pair and a document, we concatenated the two member phrases in the pair together. Table~\ref{table:data_stats} summarizes the statistics of the constructed bipartite graphs for the two datasets.

\begin{figure}[t]
\centering
\vspace{-0.4cm}
\subfigure[{Kappa coefficient}]{
\includegraphics[width = 39.5 mm]{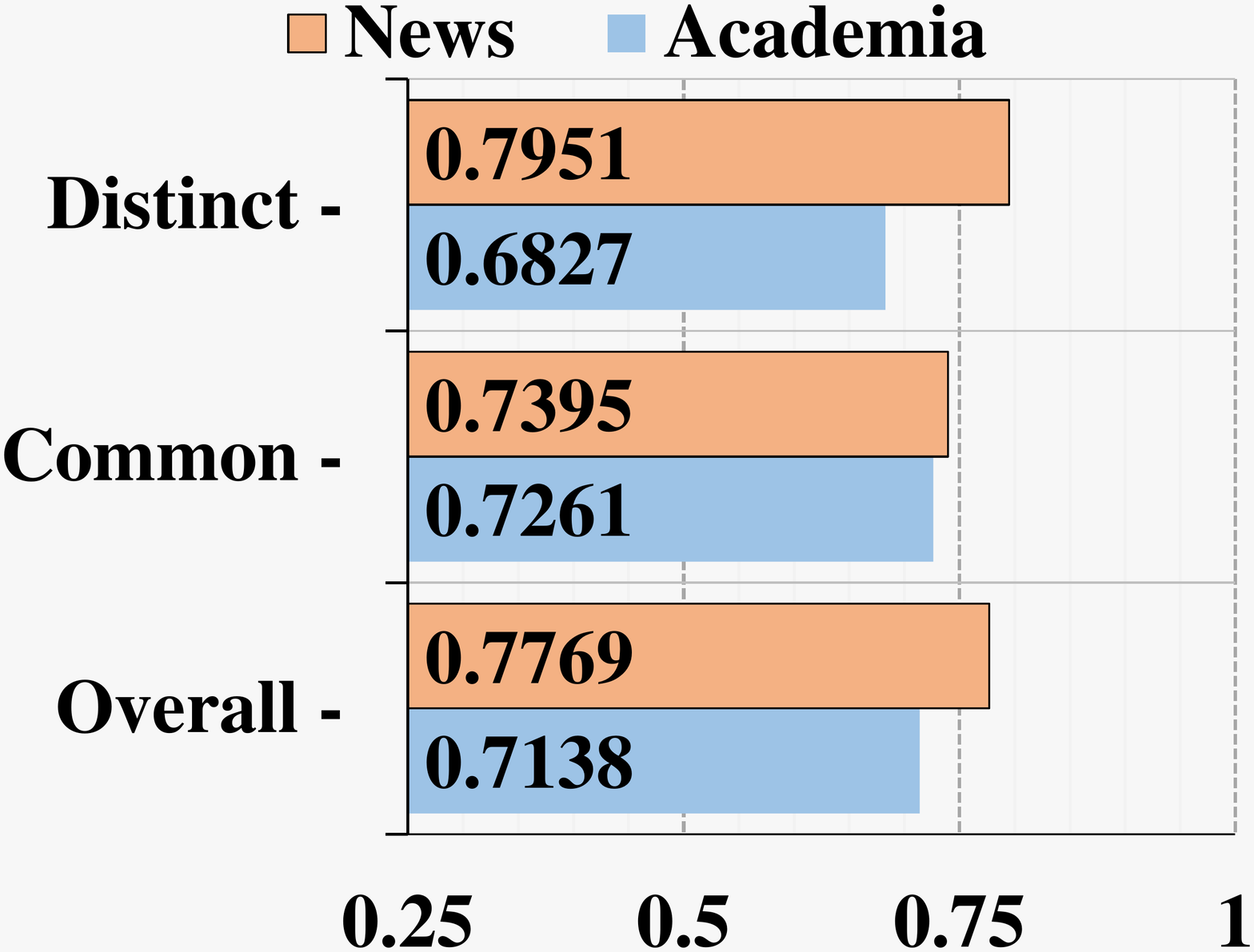}
}
\subfigure[{Relative observed agreement}]{
\includegraphics[width = 39.5 mm]{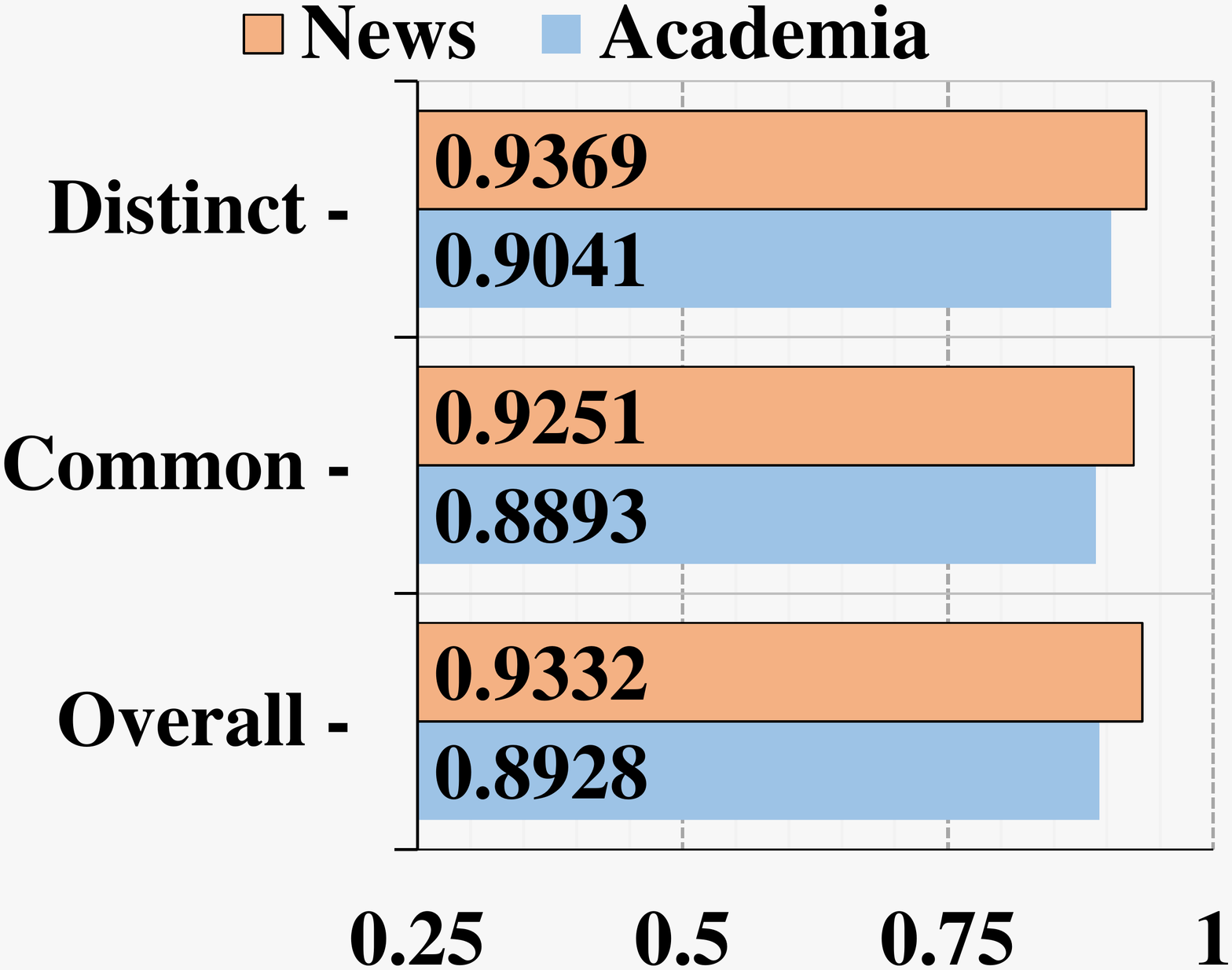}
}
\vspace{-0.2cm}
\caption{Inter-annotator agreement.}
\label{figure:inter_agreement}
\vspace{-0.2cm}
\end{figure}

\begin{table*}
\begin{scriptsize}
\vspace{-0.4cm}
\vspace{0.0cm}
\begin{center}
\begin{tabularx}{\textwidth}{  l | aaa | aaa | aaa | aaa }
\hline
 & \multicolumn{3}{c|}{\textbf{SIGIR (common)}} &  \multicolumn{3}{c|}{\textbf{SIGIR (distinct)}} & \multicolumn{3}{c|}{\textbf{KDD (common)}} & \multicolumn{3}{c}{\textbf{KDD (distinct)}}\\
\textbf{Method}
& \textbf{P} & \textbf{R}  & \textbf{F1}
& \textbf{P} & \textbf{R} & \textbf{F1}
& \textbf{P} & \textbf{R}  & \textbf{F1}
& \textbf{P} & \textbf{R}  & \textbf{F1} \\ \hline
WordMatch~\cite{mani1997summarizing} 
& 0.093 & 0.052 & 0.062
& 0.063 & 0.132 & 0.081
& 0.016 & 0.005 & 0.007
& 0.035 & 0.095 & 0.049 \\
TopicLabel~\cite{mei2007automatic}
& 0.020 & 0.016 & 0.018
& 0.158 & 0.427 & 0.226
& 0.010 & 0.003 & 0.004
& 0.100 & 0.248 & 0.137 \\
PatentCom~\cite{zhang2015patentcom} 
& 0.493 & 0.292 & 0.346
& 0.285 & \textbf{0.696} & 0.413
& 0.563 & 0.379 & 0.423
& 0.291 & \textbf{0.732} & 0.420 \\
StringFuzzy
& 0.181 & 0.815 & 0.283
& 0.261 & 0.494 & 0.329
& 0.220 & 0.802 & 0.320
& 0.299 & 0.631 & 0.391 \\
ContextFuzzy
& 0.128 & 0.839 & 0.210
& 0.259 & 0.335 & 0.281
& 0.171 & 0.796 & 0.248
& 0.301 & 0.485 & 0.338 \\ \hline
CDA-AlterMea
& 0.106 & 0.360 & 0.157
& 0.100 & 0.037 & 0.049
& 0.088 & 0.324 & 0.131
& 0.128 & 0.029 & 0.044 \\
CDA-NoGraph
& 0.628 & 0.637 & 0.630
& 0.670 & 0.529 & 0.562
& 0.799 & 0.705 & 0.716
& 0.756 & 0.645 & 0.676 \\
CDA-TwoStep
& 0.711 & 0.818 & 0.749
& \textbf{0.708} & 0.550 & 0.597
& 0.721 & 0.825 & 0.749
& 0.763 & 0.720 & 0.726 \\
CDA
& \textbf{0.752} & \textbf{0.843} & \textbf{0.778}
& 0.704 & 0.596 & \textbf{0.644}
& \textbf{0.807 }& \textbf{0.834} & \textbf{0.813}
& \textbf{0.788} & 0.711 & \textbf{0.733} \\ \hline
\end{tabularx}
\vspace{-0.2cm}
\caption{Performance comparisons on Academia dataset in terms of Precision, Recall and F1 score.}
\label{table:performance_comparison_Academia}
\end{center}
\vspace{-0.4cm}
\end{scriptsize}
\end{table*}

\smallskip
\noindent
\textsf{\textbf{3. Evaluation Sets.}}
For evaluation purposes, we selected papers published in two different areas (\ie, KDD and SIGIR) along with the News articles on Japan Tsunami to construct three evaluation sets. 
To generate document pairs \begin{small}$\U$\end{small}, we computed the document cosine similarity using TF-IDF vectors. We sampled 35 pairs of more related documents (score $>0.6$), 35 pairs of less related documents ($0.05<$ score $<0.2$) and 35 random pairs for each evaluation set.
As the number of phrases in each document is large, we adopted the \textit{pooling} method~\cite{manning2008introduction} to generate gold standard phrases. For each document pair, we constructed the pool with the results returned by all the compared methods and the top-15 salient words and phrases by WordMatch and PhraseCom. Human assessment was conducted by computer science researchers and students following the criterion in Fig.~\ref{figure:annotation_criterion}. This yields 8,463, 8,547 and 5,985 annotated phrases for KDD, SIGIR and News evaluation sets, respectively. In Fig.~\ref{figure:inter_agreement}, the kappa value and relative observed agreement both demonstrate that the human annotators have good inter-judge agreement on both common and distinct phrases.

\vspace{-0.0cm}
\subsection{Experimental Settings}
\label{subsec:experiment_setting}
In our testing of PhraseCom and its variants, we set \begin{small}$\{\lambda,\alpha\} =\{0.1, 100\}$\end{small} based on the required condition and effectiveness study on a validation set. Empirically, our performance does not change dramatically across a wide choices of parameters. For convergence criterion, we stop the outer (inner) loops in Algorithm~\ref{algorithm:PhraseCom} if the relative change of \begin{small}$\O$\end{small} in Eq.~\eqref{eq:commonality_optimization} and \begin{small}$\F$\end{small} in \eqref{eq:distinction_optimization} \big(reconstruction error in \begin{small}Eq.~\eqref{eq:relevance_score_propagation}\end{small}\big) is smaller than \begin{small}$10^{-4}$\end{small}.

\smallskip
\noindent \textsf{\textbf{Compared Methods:}}
We compared the proposed method (PhraseCom) with its variants which only model part of the proposed hypotheses. Several state-of-the-art comparative summarization methods were also implemented (parameters were first tuned on our validation sets):
(1) \textbf{WordMatch}~\cite{mani1997summarizing}: extracts top-20 \textit{salient words} based on TF-IDF scores. It generates common set based on (a) in Fig.~\ref{figure:proposed_measures} and takes the rest as distinct sets;
(2)
\textbf{TopicLabel}~\cite{mei2007automatic}: TopicLabel selects salient phrases based on first-order relevance measure and uses same method as WordMatch to generate common and distinct sets;
(3)
\textbf{PatentCom}~\cite{zhang2015patentcom}: a state-of-the-art graph-based comparative summarization method. We adopt its common and distinct phrase sets for comparisons;
(4)
\textbf{StringFuzzy}: follows Sec.~\ref{subsec:candidate_generation} to extract salient phrases. It finds common phrase if its BM25 scores to \textit{both} documents are larger than a threshold (3.0) and uses the remaining salient phrases to form distinct sets;
and (5)
\textbf{ContextFuzzy}: Different from StringFuzzy, it measures cosine similarity between the pseudo-document of a phrase (all contexts in a 10 words window in the corpus) and a document.

For PhraseCom, besides the proposed full-fledged model, \textbf{CDA}, we also compare with its variants which implement our intuitions differently:
(1)
\textbf{CDA-NoGraph}: It uses BM25 scores to compute commonality and distinction measures and optimizes Eqs.~\eqref{eq:commonality_optimization} and \eqref{eq:distinction_optimization} without graph-based regularization \begin{small}$\L$\end{small};
(2)
\textbf{CDA-AlterMea}: It uses summation of relevance scores as a commonality measure and differences between relevance scores as a distinction measure;
and (3)
\textbf{CDA-TwoStep}: It first learns relevance score based on Eq.~\eqref{eq:relevance_score_propagation} and then optimizes Eqs.~\eqref{eq:commonality_optimization} and \eqref{eq:distinction_optimization} without \begin{small}$\L$\end{small}.

\smallskip
\noindent \textsf{\textbf{Evaluation Metrics:}}
We use F1 score computed from Precision and Recall to evaluate the performance. Given a document pair, we denote the set of system-identified common terms as \begin{small}$\I$\end{small} and the set of gold standard common terms (\ie, words and phrases which are judged as good or perfect) as \begin{small}$\G$\end{small}. Precision (P) is calculated by \begin{small}$\text{P}=|\I\cap\G|/|\I|$\end{small} and Recall (R) is calculated by \begin{small}$\text{R}=|\I\cap\G|/|\G|$\end{small}. For each document in the pair, we compute above metrics for distinct terms in a similar manner. The reported numbers are averaged over the evaluation set. For parameter study in validation set, we use the same metrics to evaluate the performance.

\begin{table}
\begin{scriptsize}
\vspace{-0.0cm}
\vspace{0.0cm}
\begin{center}
\begin{tabularx}{\linewidth}{  l | xxx | xxx } \hline
& \multicolumn{3}{c|}{\textbf{Common}}
& \multicolumn{3}{c}{\textbf{Distinct}} \\
\textbf{Method}
& \textbf{P} & \textbf{R}  & \textbf{F1}
& \textbf{P} & \textbf{R} & \textbf{F1}\\ \hline
WordMatch~\cite{mani1997summarizing} 
& 0.035 & 0.064 & 0.045
& 0.079 & 0.221 & 0.112 \\
TopicLabel~\cite{mei2007automatic}
& 0.327 & 0.449 & 0.363
& 0.412 & 0.851 & 0.534 \\
PatentCom~\cite{zhang2015patentcom} 
& 0.358 & 0.481 & 0.399
& 0.434 & \textbf{0.877} & 0.554 \\
StringFuzzy
& 0.180 & 0.414 & 0.245
& 0.376 & 0.735 & 0.470 \\
ContextFuzzy
& 0.166 & 0.422 & 0.222
& 0.317 & 0.661 & 0.397 \\ \hline
CDA-AlterMea
& 0.347 & 0.613 & 0.399
& 0.264 & 0.194 & 0.215 \\
CDA-NoGraph
& 0.600 & 0.488 & 0.521
& 0.838 & 0.687 & 0.727 \\
CDA-TwoStep
& 0.642 & 0.840 &  0.639
& 0.831 & 0.726 & 0.753 \\
CDA
& \textbf{0.704} & \textbf{0.878} & \textbf{0.757}
& \textbf{0.871} & 0.723 & \textbf{0.773} \\\hline
\end{tabularx}
\vspace{-0.2cm}
\caption{Performance comparisons on News dataset in terms of Precision, Recall and F1 score.}
\label{table:performance_comparison_News}
\end{center}
\vspace{-0.4cm}
\end{scriptsize}
\end{table}

\begin{figure*}
\centering
\vspace{-0.4cm}
\includegraphics[width = 158mm]{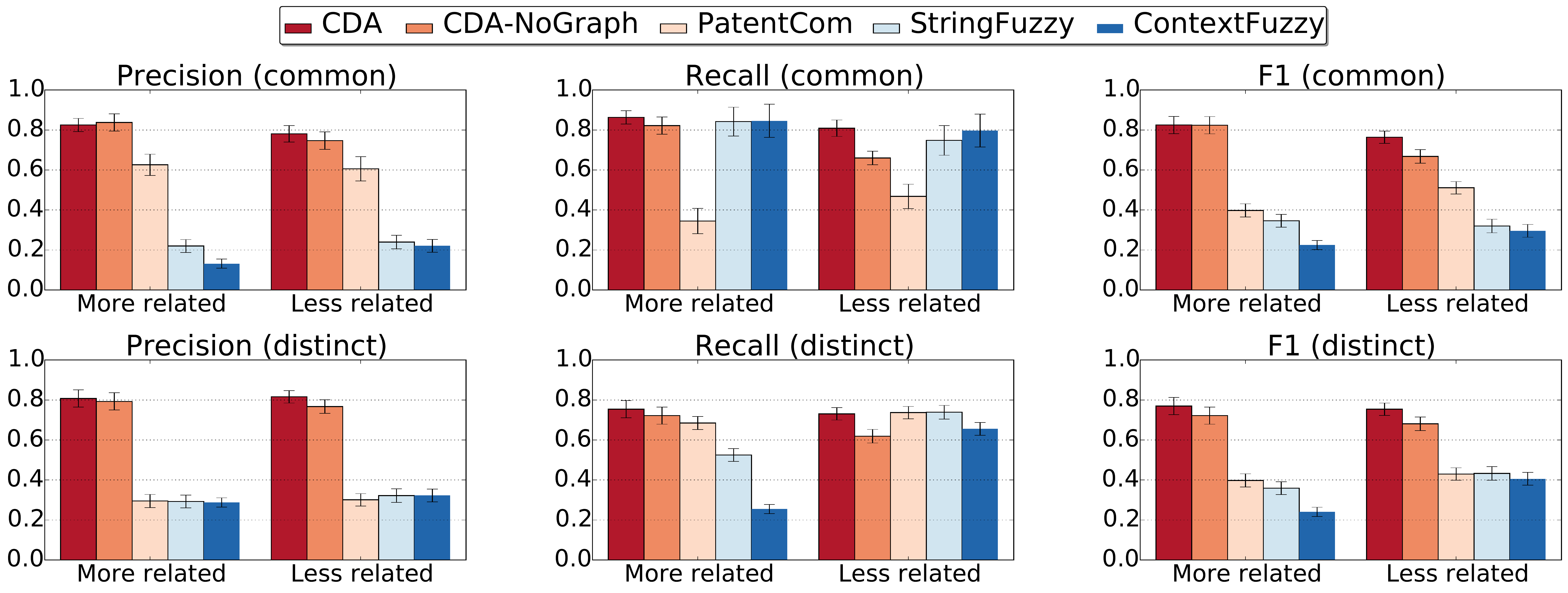}
\vspace{-0.2cm}
\caption{Performance study on ``more related" and ``less related" document pairs on the Academia dataset.}
\label{figure:performance_study_different_pairs}
\vspace{-0.1cm}
\end{figure*}

\vspace{-0.0cm}
\subsection{Experiments and Performance Study}
\label{subsec:performance_comparison}

\noindent
\textsf{\textbf{1. Comparing CDA with the other methods.}}
Tables~\ref{table:performance_comparison_Academia} and \ref{table:performance_comparison_News} summarize the comparison results on the three evaluation sets. 
Overall, CDA outperforms others on all metrics on all evaluation sets in terms of finding commonalities, and achieves superior Precision and F1 scores on finding distinctions.
In particular, CDA obtains a 125\% improvement in F1 score and 188\% improvement in Recall on the SIGIR dataset compared to the best baseline PatentCom on finding commonalities, and improves F1 on the News dataset by 39.53\% compared to PatentCom on finding distinctions.

PatentCom uses feature selection technique to jointly discover distinct words and phrases for a document pair but suffers from low recall on commonalities since it finds common terms simply by term overlap without considering semantic common words/phrases (same as WordMatch and TopicLabel). 
Although its recall on distinctions is high, it has low precision, since it returns a large number of distinct terms without filtering those overly specific ones. Superior performance of CDA validates the effectiveness of our salient phrase generation (vs. WordMatch and TopicLabel) and of the proposed hypotheses on modeling semantic commonality and document-phrase relevance.
Both StringFuzzy and ContextFuzzy can find semantic common phrases but they suffer from low precision and instable recall (49.7\% drop from SIGIR to News) due to its sensitivity to the cut-off threshold. A one-fit-all threshold is not guaranteed to work well for different document pairs or for different domains. It is worth mentioning that CDA performs more stably because it leverages the graph-based semantic relevance and the constraints in the optimization, which can control the size of the output sets automatically like an adaptive thresholding algorithm.

\smallskip
\noindent
\textsf{\textbf{2. Comparing CDA with its variants.}}
Comparing with CDA-NoGraph and CDA-TwoStep, CDA gains performance from propagating semantic relevance on graphs and integrating relevance propagation with phrase selection in a mutually enhancing way. It dramatically outperforms CDA-AlterMea on all metrics on all evaluation sets, which demonstrates the effectiveness of the proposed commonality and distinction measures (see Sec.~\ref{subsec:model}). 

\smallskip
\noindent
\textsf{\textbf{3. More related pairs versus less related pairs.}}
Fig.~\ref{figure:performance_study_different_pairs} compares the methods on pairs of highly similar (more related) documents and pairs of lowly similar (less related) documents, respectively.
CDA and its variant outperform other methods in terms of Precision and F1 scores on both kinds of document pairs. As there exist more semantic commonalities and subtle differences between a pair of highly similar documents, CDA gains larger improvement by optimizing the proposed measures and by learning semantic relevance.
The superior Recall of CDA over CDA-NoGraph (a 23\% improvement on the less related pairs) mainly comes from the graph-based relevance propagation.

\subsection{Case Study}
\noindent
\textsf{\textbf{1. Example output on Academic papers.}}
Table~\ref{table:case_study_Academia} compares the output of CDA with that of Microsoft Rich-Caption system on two sample papers. Rich-Caption system provides structured information (\eg, related concepts) for papers on Bing Academic vertical (for United States users). However, it suffers from low coverage on the identified related concepts. 
CDA not only can highlight the subtle differences between two similar papers (first example) but also can find the semantic commonalities (\eg, ``\textit{EM algorithm}") between two moderately similar papers (second example).

\begin{figure*}[t]
\centering
\vspace{-0.3cm}
\includegraphics[width=165mm]{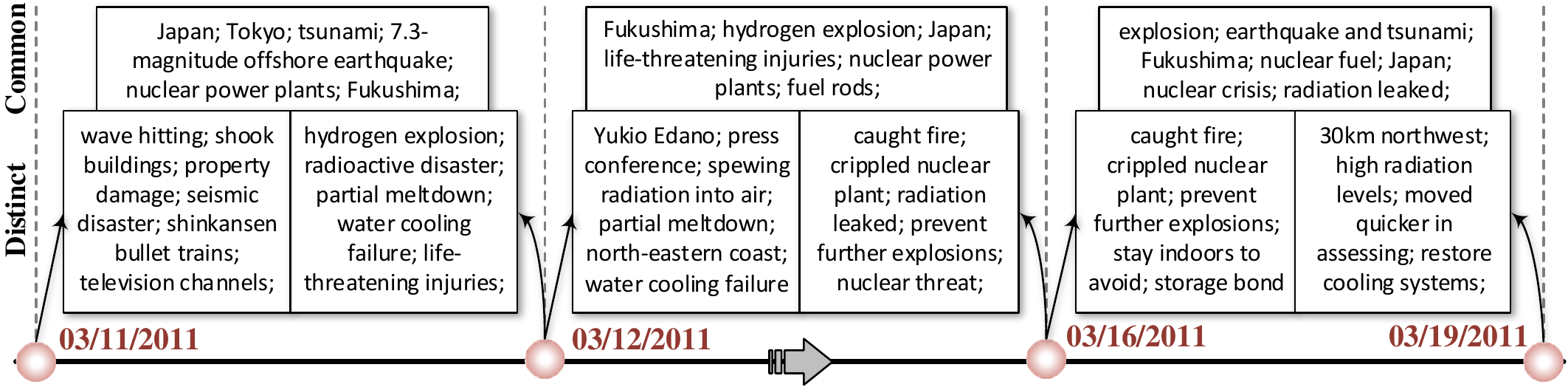}
\vspace{-0.2cm}
\caption{Compare document sets on News dataset. We use news articles published at four different dates.}
\label{figure:case_study_News}
\vspace{-0.0cm}
\end{figure*}

\smallskip
\noindent
\textsf{\textbf{2. Comparing two document sets.}}
Fig.~\ref{figure:case_study_News} presents CDA output for comparing document sets on News dataset (20 documents were sampled for each date). Common phrases show the connections between things happened in two different dates while distinction phrases highlight the unique things happened in each date. In particular, distinction results demonstrate that our method can capture pairwise distinction property well by generating different distinct phrases for the same news set when comparing with different news sets. The comparison results provide a good overview on the event evolution of 2011 Japan Tsunami.

\begin{figure}[t]
\centering
\vspace{-0.0cm}
\subfigure[{Semantic Commonality}]{
\includegraphics[width = 39.5 mm]{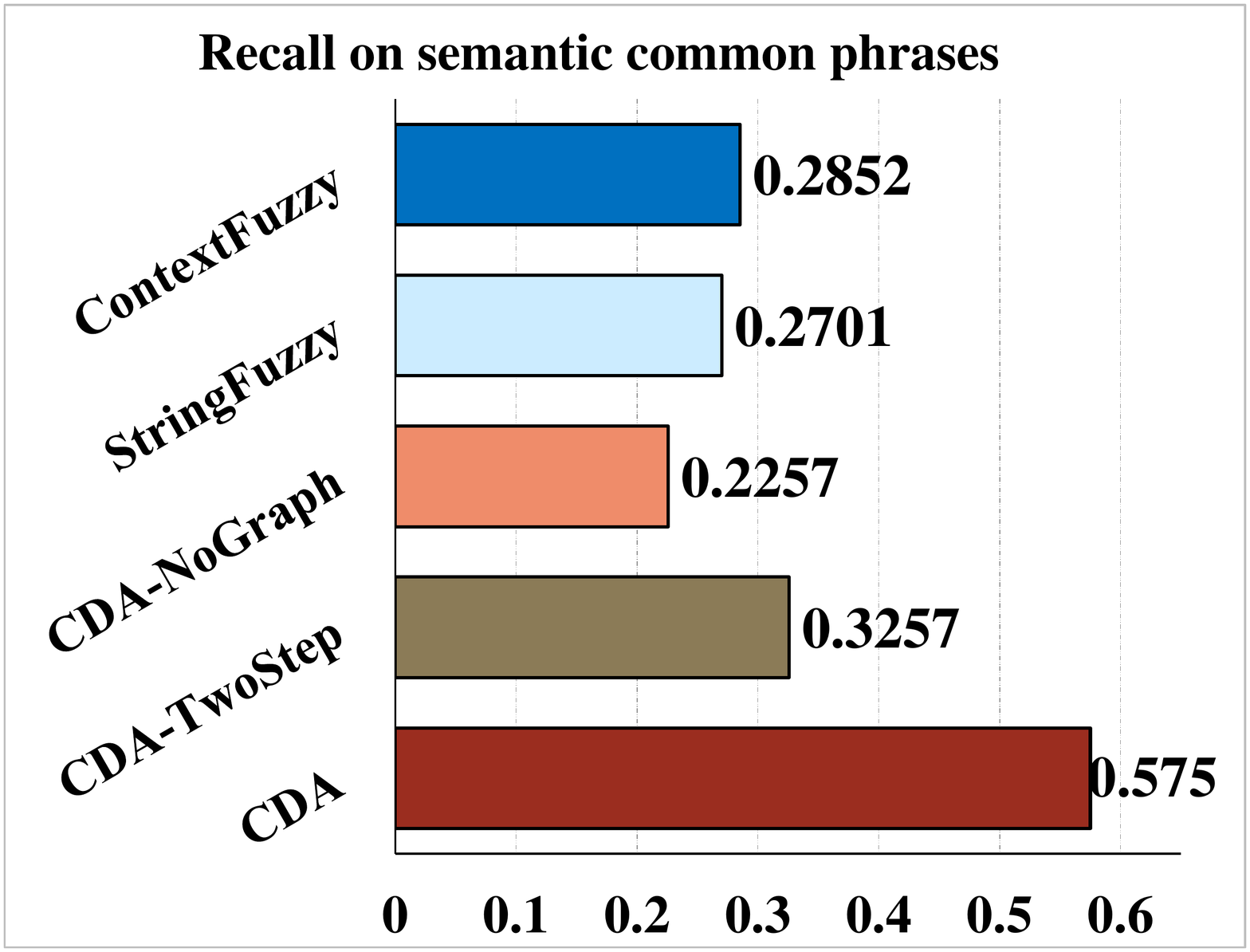}
\label{figure:semantic_commonness}
}
\subfigure[{Perfect Distinction}]{
\includegraphics[width = 39.5 mm]{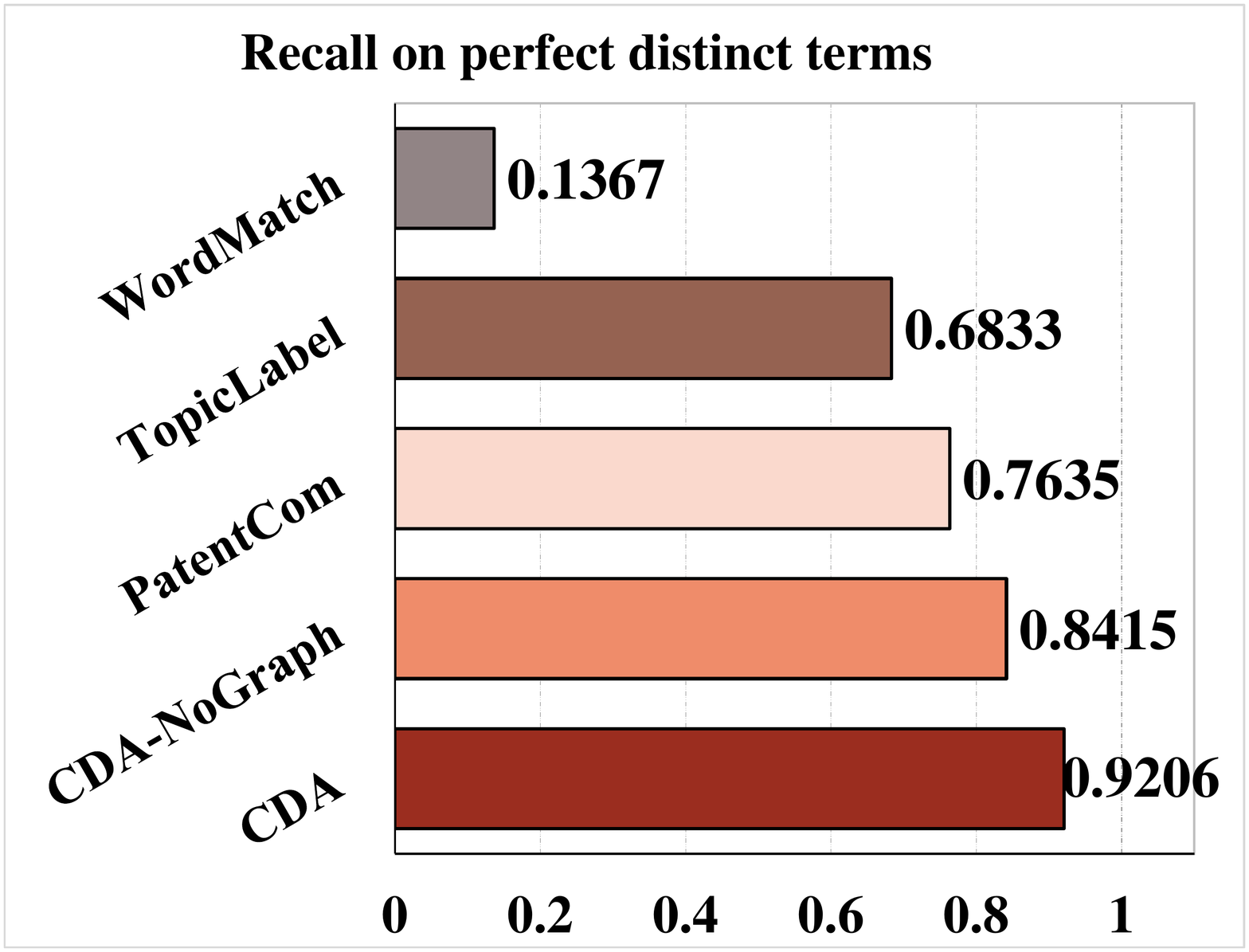}
\label{figure:perfect_distinction}
}
\vspace{-0.4cm}
\caption{Case studies on phrase semantic commonality and perfect distinction on the News dataset.}
\vspace{-0.3cm}
\end{figure}

\smallskip
\noindent
\textsf{\textbf{3. Testing on semantic commonality.}}
To study the performance on finding semantic common phrases (Fig.~\ref{figure:case_study_commonality}), we compare our method with  methods that can also find such phrases. In Fig.~\ref{figure:semantic_commonness}. CDA achieved significant improvement in recall since it leverages the bipartite graph to derive semantic relevance (versus CDA-NoGraph, StringFuzzy, ContextFuzzy), and integrates the relevance score propagation with phrase selection to reinforce the learning of semantic relevance (versus CDA-TwoStep). Compared with StringFuzzy and ContextFuzzy, our method does not require a unified cut-off threshold and thus is more robust.

\begin{table}[t]
\begin{small}
\vspace{-0.0cm}
\begin{center}
\begin{tabularx}{\linewidth}{| X | X |} \hline
\multicolumn{1}{|c|}{\textbf{Distinctions of}~\cite{jeh2003scaling}} & \multicolumn{1}{c|}{\textbf{Distinctions of}~\cite{haveliwala2003topic}} \\
\centering \textbf{Keywords}: search, Web graph, link structure, \textit{PageRank}, search in context,  personalized search & \begin{center}\textbf{Keywords}: web search, \textit{PageRank}\end{center}
\\ \hline
hub, partial, skeleton, pages, personalized, section, basis, computation, preference
& query, rank, htm, sensitive, ranking, urls, search, topic, context, regional, kids
\\ \hline
The Hubs Theorem allows basis vectors to be encoded as partial vectors and a hubs skeleton. Our approach enables incremental computation, so that the construction of personalized views from partial vectors is practical at query time. 
& Finally, we compute the query-sensitive importance score of each of these retrieved URLs as follows. In Section 4.3, the topic-sensitive ranking vectors were chosen using the topics most strongly associated with the query term contexts.
\\ \hline
personalized web search, $\newline$ user-specified@@web pages, $\newline$ dynamic programming, $\newline$ incremental computation, $\newline$ theoretical results
& topic-sensitive PageRank, $\newline$ context-specific importance  $\newline$ score,~query context, topic- $\newline$ sensitive@@  ranking vector, $\newline$ query topic
\\ 
\hline
\end{tabularx}
\vspace{-0.1cm}
\caption{Distinctions for papers \cite{jeh2003scaling} and \cite{haveliwala2003topic} generated by WordMatch~\cite{mani1997summarizing} (top), Discriminative Sentence Selection~\cite{wang2012comparative} (middle), and CDA (bottom).}
\label{table:case_study_phraseVSsentence}
\end{center}
\vspace{-0.5cm}
\end{small}
\end{table}

\begin{table*}
\begin{scriptsize}
\vspace{-0.0cm}
\vspace{0.0cm}
\begin{tabularx}{\textwidth}{ | t | X|>{\hsize=.4\hsize}X | X|>{\hsize=.4\hsize}X |}
\hline
& \multicolumn{2}{p{81mm}|}{\textbf{$d$:~Document Summarization Based on Word Associations
$\newline$ $\d'$:~Cross-Document Summarization by Concept     $\newline$ Classification} }
& \multicolumn{2}{p{81mm}|}{$d$:~\textbf{A Latent Topic Model for Linked Documents}
$\newline$ $\d'$:~\textbf{Document clustering with cluster refinement and model selection capabilities}}
\\ \hline
& \multicolumn{1}{c|}{CDA} & Microsoft system & \multicolumn{1}{c|}{CDA} & Microsoft system
\\ \hline
$\C$ 
& multi-document summarization; text summarization; document summary 
& multi-document summarization
& document clustering; clustering task@@em algorithm; expectation-maximization;
& document clustering
\\ \hline
$\Q$  
& sentence selection@@word associations; background corpus; association mixture text summarization; language-independent; unsupervised method; novelty detection;
& \multicolumn{1}{c|}{N/A}
& plsi model@@linked documents; latent semantic space; link information; citation-topic model;
& digital library; stochastic process; topic model
\\ \hline
$\Qp$
& passage similarity@@term weights; cross-document summarizer;  clustering techniques; discourse structure; relevance measure; conceptual clustering;
& n-gram; $\newline$ cluster analysis
& voting scheme@@document cluster labels; gaussian mixture model; discriminative feature set@@cluster label; majority voting@@feature set; 
& expectation maximization algorithm; mixture model; model selection
\\ \hline
\end{tabularx}
\vspace{-0.0cm}
\caption{Example output of CDA and Microsoft Academic Rich-Caption system on two papers from the Academia Dataset.}
\label{table:case_study_Academia}
\vspace{-0.2cm}
\end{scriptsize}
\end{table*}

\smallskip
\noindent
\textsf{\textbf{4. Testing on perfect distinction.}}
We consider \textit{overly general} terms as positive in previous evaluations (\ie, ``good" labels are given to overly general terms in Sec.~\ref{subsec:experiment_setting}). Next, we further test our method particularly on finding ``perfect" distinct terms (by assigning ``bad" label to those overly general terms). 
In Fig.~\ref{figure:perfect_distinction}, CDA achieved superior performance (over 90\% on recall) compared with other methods. This is because that (1) phrase is not only informative and concise enough for user to understand, but also general enough to highlight the distinctions (versus WordMatch); and (2) our relevance score learning can balance phrase generality and discrimination well so as to be not dominated by overly general terms. (versus CDA-Graph, PatentCom, TopicLabel).

\smallskip
\noindent
\textsf{\textbf{5. Comparing with word-based and sentence-based summarization.}}
Table~\ref{table:case_study_phraseVSsentence} shows the comparative analysis results between papers \cite{jeh2003scaling} and \cite{haveliwala2003topic} generated by word-based method~\cite{mani1997summarizing}, sentence-based method~\cite{wang2012comparative} (top-2 sentences) and our phrase-based method. We do not include commonality results since sentence-based summarization techniques only provide distinction results~\cite{wang2012comparative} for each document.
We found that CDA provides sufficiently cohesive and readable results compared with word-based methods, and it keeps the overall summary concise, as compared to sentence-based methods.
Furthermore, the results also show that author-generated keywords are often not able to highlight the distinctions when compared to other papers.

\section{Related Work}
\label{sec:related}
There have been many attempts on performing comparative analysis on text data. Previous work can be categorized in terms of sources of comparison (\eg, single or multiple text collections), targets of comparison (\eg, between topics, individual documents or document sets), aspects to compare (\eg, commonality, distinction or both), and representation forms of results (\eg, words, sentences).

Multi-document summarization~\cite{shen2010multi,haghighi2009exploring,lin2002single,conroy2001text,radev2000centroid} aims to generate a compressed summary to cover the consensus of information among the original documents. It is different from \textit{commonality} discovery as it focuses on providing comprehensive view of the corpus (union instead of intersection).

Comparative document summarization~\cite{zhang2015patentcom,arun15emnlp,huang2014comparative,wang2012comparative,zhuang2012mining,huang2011comparative} focuses on the \textit{distinction} aspect---it summarizes the differences between comparable document sets by extracting the discriminative sentences from each set. Existing work formalizes the problem of discriminative sentence selection into different forms, including integer linear programming~\cite{huang2014comparative}, multivariate normal model estimation~\cite{wang2012comparative}, and group-related centrality estimation~\cite{mani1997summarizing}.
Going beyond selecting discriminative sentences from a document set, our problem aims to select quality and concise phrases to highlight not only differences but also commonalities between two documents.

In particular, our work is related to \cite{mani1997summarizing} since both try to derive common and distinct terms for a pair of related documents, but their work focuses on finding topic-related terms and selecting sentences based on such terms. They assume terms appearing in both documents as the common ones, and treat the remaining terms as distinct ones. As shown in our experiments, this method (labeled as WordMatch) suffers from low recall in finding common terms and low precision in finding distinct terms since it ignores semantic common phrases and pairwise distinct phrases.
Similarly, Zhang~\etal~\cite{zhang2015patentcom} consider both common and distinct aspects in generating comparative summary for patent documents. They derive a term co-occurrence tree which can be used to extract sentences for summarization. However, they use all shared noun phrases between two documents as common terms, and apply feature selection techniques to find distinct terms. This method (see PatentCom in Sec.~\ref{subsec:performance_comparison}) demonstrates poor performance on finding common terms due to the ignorance of semantic common phrases; although this method performs well in terms of the recall for distinct phrases, it achieves low precision, since it fails to model phrase generality and produce many overly-specific phrases. 

Another line of related work, referred to as comparative text mining~\cite{zhai2004cross}, focuses on modeling latent comparison aspects and discovering common and collection-specific word clusters for each aspect. 
They adopt topic model~\cite{chen2015differential,zhai2004cross,lu2008opinion} and matrix factorization~\cite{kim2015nmf} to present the common and distinct information by multinomial distributions of words.
While latent comparison aspects can help enrich the comparison results, these methods still adopt bag-of-words representation which is often criticized for its unsatisfying readability~\cite{mei2007automatic}. Furthermore, it is difficult to apply statistical topic modeling in comparative document analysis as the data statistic between two documents is insufficient, in particular when the documents are about emergent topics (\eg, Academia, News).
Finally, our work is also related to comparing reviews in opinion mining~\cite{sipos2013generating,paul2010summarizing,kim2009generating,jindal2006identifying} and contrastive summarization for entities~\cite{lerman2009contrastive}---they also aim to find similarities and differences between two objects. However, these works are restricted to sentiment analysis.

\section{Conclusion and Future Work}
\label{sec:conclusion}
In this paper, we study the problem of phrase-based comparative summarization for a document pair, called comparative document analysis (CDA), and propose a general graph-based approach to model semantic commonality and pairwise distinction for phrases.
We cast the phrase selection problems into joint optimization problems based on the proposed novel measures.
Experiment results demonstrate the effectiveness and robustness of the proposed method on text corpora of different domains. 
Interesting future work includes extending CDA to consider different comparison aspects~\cite{zhai2004cross,kim2015nmf} and to exploit the hierarchical semantic relations between phrases. CDA is general and can be applied as a primitive step for sentence-based comparative summarization~\cite{huang2014comparative,wang2012comparative}. It can potentially benefit many other text applications such as content recommendation, document clustering and relevance feedback in information retrieval~\cite{manning2008introduction}.

\newpage
\bibliographystyle{abbrv}
\bibliography{15-cda}

\end{document}